\documentclass{aa}
\usepackage{txfonts}
\usepackage{natbib,twoopt}
\usepackage{longtable}
\usepackage{pdflscape}
\usepackage{array}
\usepackage{graphicx}
\usepackage{threeparttable}
\usepackage{setspace}
\usepackage{arydshln}
\usepackage{enumitem}
\usepackage{multirow}
\usepackage{multicol}
\usepackage{hyperref}
\usepackage{url}
\bibpunct{(}{)}{;}{a}{}{,} 

\hypersetup{
	pdftitle={Formation of the G-band in 3D},
	pdfauthor={A. J. Gallagher},
	pdfsubject={A\&A journal},
	pdfcreator={A. J. Gallagher},
		colorlinks, 
		citecolor=blue, 
		filecolor=blue, 
		linkcolor=blue,
		breaklinks=true, 
		plainpages=false,
		urlcolor=blue 
}

\title{An in-depth spectroscopic examination of molecular bands from 3D hydrodynamical model atmospheres}
\subtitle{I. Formation of the G-band in metal-poor dwarf stars}

	\author{A. J. Gallagher\inst{1}\thanks{Observatoire de Paris fellow}
\and
E. Caffau\inst{1}\thanks{MERAC fellow}
\and
P. Bonifacio\inst{1}
\and 
H.-G. Ludwig\inst{2,1}
\and
M. Steffen\inst{3,1}
\and
M. Spite\inst{1}
}

\institute{GEPI, Observatoire de Paris, PSL Research University, CNRS, Univ. Paris Diderot, 
Sorbonne Paris Cit\'{e} Place Jules Janssen, 92190 Meudon, France.\\ 
email: \texttt{andrew.gallagher@obspm.fr}
\and Zentrum f{\"u}r Astrononmie der Universit{\"a}t Heidelberg, Landessternwarte, K{\"o}nigstuhl 12, 69117 Heidelberg, Germany.
\and Leibniz-Institut f{\"u}r Astrophysik Potsdam, An der Sternwarte 16, 14482 Potsdam, Germany.
}

\date{Received 29 March 2016 / Accepted 20 May 2016}

\authorrunning{A. J. Gallagher et al.}
\titlerunning{Formation of the G-band in 3D}

\newcommand{\cobold}{CO$^{5}$BOLD}
\newcommand{\atd}{$\langle{\rm 3D}\rangle$}
\newcommand{\odx}{{\tt LHD}}
\newcommand{\teff}{T_{\rm eff}}
\newcommand{\logg}{\log{g}}
\newcommand{\loggf}{\log{gf}}
\newcommand{\kms}{{\rm km\,s^{-1}}}
\newcommand{\feh}{{\rm [Fe/H]}}
\newcommand{\logtr}{\log{\tau_{\rm ROSS}}}
\newcommand{\ctd}{\Delta_{\rm 3D - 1D}}
\newcommand{\actd}{A({\rm C})_{\rm 3D}}
\newcommand{\aotd}{A({\rm O})_{\rm 3D}}
\newcommand{\acod}{A({\rm C})_{{\rm 1D},\odx}}
\newcommand{\aood}{A({\rm O})_{{\rm 1D},\odx}}

\abstract
{
Recent developments in the three-dimensional (3D) spectral synthesis code Linfor3D have meant that, for the first time, large spectral wavelength regions, such as molecular bands, can be synthesised with it in a short amount of time.
}
{
A detailed spectral analysis of the \emph{synthetic} G-band for several dwarf turn-off-type 3D atmospheres ($5850\lesssim\teff\,[{\rm K}]\lesssim6550$, $4.0\leq\logg\leq4.5$, $-3.0\leq\feh\leq-1.0$) was conducted, under the assumption of local thermodynamic equilibrium. We also examine carbon and oxygen molecule formation at various metallicity regimes and discuss the impact it has on the G-band.
}
{
Using a qualitative approach, we describe the different behaviours between the 3D atmospheres and the traditional one-dimensional (1D) atmospheres and how the different physics involved inevitably leads to abundance corrections, which differ over varying metallicities. Spectra computed in 1D were fit to every 3D spectrum to determine the 3D abundance correction. 
}
{
Early analysis revealed that the CH molecules that make up the G-band exhibited an oxygen abundance dependency; a higher oxygen abundance leads to weaker CH features. Nitrogen abundances showed zero impact to CH formation. The 3D corrections are also stronger at lower metallicity. Analysis of the 3D corrections to the G-band allows us to assign estimations of the 3D abundance correction to most dwarf stars presented in the literature.
}
{
The 3D corrections suggest that $A({\rm C})$ in CEMP stars with high $A({\rm C})$ would remain unchanged, but would decrease in CEMP stars with lower $A({\rm C})$. It was found that the C/O ratio is an important parameter to the G-band in 3D. Additional testing confirmed that the C/O ratio is an equally important parameter for OH transitions under 3D. This presents a clear interrelation between the carbon and oxygen abundances in 3D atmospheres through their molecular species, which is not seen in 1D.
}

\begin{document}

\maketitle

\section{Introduction}
\label{sec:introduction}

The acquisition and analysis of very metal-poor ($\feh$\footnote{$\left[{\rm X/Y}\right]=\log_{10}{\left(\dfrac{N({\rm X})}{N({\rm Y})}\right)}_* - \log_{10}{\left(\dfrac{N({\rm X})}{N({\rm Y})}\right)}_\sun$}$<-2.0$) stellar spectra allows us to unravel the chemical history of the Galaxy and helps us understand the chemical processes of the early Universe and how they might differ from those observed today. A considerable fraction \citep[$\sim20\%$,][]{Lucatello2006} of all metal-poor stars are known to be enhanced in carbon \citep[${\rm [C/Fe]}>+1.0$,][]{Beers2005}. We refer to these chemically peculiar stars as carbon-enhanced metal-poor (CEMP) stars. Slightly different definitions of CEMP stars have been proposed \citep[see e.g.][and references therein]{Hansen2016}, but we prefer to keep the ``classical'' definition, for the reasons explained in \citep{Bonifacio2015}. As CEMP stars make up a considerable fraction of metal-poor stars, it is important to understand how they acquired their enhanced carbon abundance.

CEMP stars can be divided into several sub-classes, according to their heavy-element abundances. Heavy elements (i.e. $Z\ge 30$) mostly form via two neutron-capture processes; the slow (s-) process and the rapid (r-) process. Some heavy elements, such as barium, are formed mostly via the s-process path and others, such as europium, mostly via the r-process path \citep{Arlandini1999}. CEMP-s stars are defined as ${\rm [C/Fe]}>+1.0$, ${\rm [Ba/Fe]}>+1.0$ \& ${\rm [Ba/Eu]>+0.5}$ \citep{Beers2005}, i.e. elements that form along the s-process path are more abundant that those that form along the r-process path. CEMP-r/s stars are defined as ${\rm [C/Fe]}>+1.0$ \& $0.0<{\rm [Ba/Eu]<+0.5}$ \citep{Beers2005}, indicating that the abundances of s-process material is roughly equal to the abundances of r-process material. CEMP-s and CEMP-r/s stars make up $\sim80\%$ of all CEMP stars known \citep{Aoki2007,Aoki2008}. In general, CEMP stars with $\feh<-3.4$ are CEMP-no stars \citep{Aoki2007}, i.e. ${\rm [C/Fe]}>+1.0$, ${\rm [Ba/Fe]}<0.0$, \citep{Beers2005}. These stars are of particular interest as almost all of them show no indication of binarity, suggesting that their carbon enhancement is indicative of the gas cloud from which they formed. Additionally, when the metallicity is further reduced to $\feh<-4.5$, all but one \citep[i.e. SDSS J1029+1729,][]{Caffau2011} of them is a CEMP star, including the most iron-poor star currently known, SMSS\,J031300.36--670839.3 \citep{Keller2014}, where recent measurements of the carbon abundance found that ${\rm [C/Fe]}> +5.11$ \citep{Bessell2015}. This suggests that carbon-enhancement has some unknown role in the chemical processes of the early Universe. However, these extremely metal-poor stars are very rare. Only nine stars within this metallicity regime are currently known of \citep{Bonifacio2015,Frebel2015,FN2015}.

Measurement of a star's carbon abundance can be done several ways. However, the most common measurement utilises a molecular feature of CH, the G-band ($A^{2}\Delta-X^{2}\Pi$). This requires the computation of the emerging flux in the G-band from a model atmosphere. Over the past decade, considerable efforts have been made to improve the modelling process.  Three-dimensional (3D) dynamic model atmospheres realistically simulate the convective motions of a turbulent stellar atmosphere, removing the necessity for common ``fudge'' parameters, such as the mixing length parameter ($\alpha_{\rm MLT}$), macroturbulence and microturbulence. Their production is very computationally expensive, relative to the more simplistic 1D alternative, however. Also, synthesising large wavelength regions from a 3D model atmosphere is very difficult to achieve \citep[see e.g.][]{Bonifacio2013}. \citet{Frebel2008} tackled the problem by introducing corrections to the {\it f}-values of the molecular transitions in their 1D synthesis. This technique is interesting, yet difficult to generalise (a new set of ``corrected'' {\it f}-values has to be computed for each set of atmospheric parameters) and it ignores all effects of velocity fields (line shifts, turbulence on all scales). It is thus not surprising that it has not been further pursued.

This paper represents the first in a series papers on molecular bands in 3D. In it we present, for the first time, an extensive look at the formation of the G-band in a 3D atmosphere and present \emph{synthetic} 3D abundance corrections to the G-band. Until very recently such a calculation was not computationally possible, but advances in the way the 3D spectrum synthesis code Linfor3D \citep{Steffen2014} handles the computation of large wavelength regions now means that such work can be done.

This paper is organised as follows: Sect.~\ref{sec:modelling_gband} presents the 3D model atmospheres, line lists and spectrum synthesis in detail; Sect.~\ref{sec:gbandformation} details the formation of the G-band in 3D and examines how the different physics involved in 1D and 3D spectrum synthesis and model atmosphere computation affect it; Sect.~\ref{sec:gband_analysis} presents the 3D abundance corrections to the G-band; Sect.~\ref{sec:discussion} discusses further implications of our findings; and Sect.~\ref{sec:conclusions} presents the conclusions drawn from the work.

\section{Modelling the G-band in 3D}
\label{sec:modelling_gband}

The G-band was computed under the assumption of LTE using the 3D spectrum synthesis code Linfor3D \citep{Steffen2014}, which utilises 3D hydrodynamic LTE model atmospheres produced by \cobold\ \citep{Freytag2012}.

\subsection{The 3D model atmospheres}
\label{sec:modelatmospheres}

A typical 3D atmosphere consists of a series of temporal structures or computational boxes, usually referred to as snapshots. They are selected from a larger series of snapshots, output by \cobold, taken from simulation sequences that have reached full dynamical and thermal relaxation. Typically, there are 20 selected snapshots for each \cobold\ atmosphere that are used for spectral synthesis, though there are those with less. All the models used in this paper are computed using the box-in-a-star mode in \cobold, i.e. the computational box is a small part of the atmosphere. We apply the ergodic hypothesis, thus assuming that averaging over time (different snapshots) is equivalent to averaging over space (the whole stellar disc). Thus we assume our ``flux'' spectra averaged over snapshots can be directly compared to observed spectra, that are integrated over the whole stellar disc.

The geometrical size of the computational box is dependent on the model's temperature and gravity. It is scaled by the solar model ($5.6\times5.6\times2.3\,{\rm Mm}$) according to the resultant pressure scale height at the surface. This means that models with the same temperature and gravity but different metallicities can be directly compared without the need to rescale the computational box. At present, there are a large number of precomputed \cobold\ model atmospheres available through the ``Cosmological Impact of the First STars'' (CIFIST) collaboration \citep{Ludwig2009}. As they are computed using the scaling described above, the resultant synthesis of two CIFIST models of different metallicity (but equal temperatures and gravities) can be interpolated to attain a rescaled spectrum with a metallicity, not currently available in the CIFIST grid, as was done in \citet{Gallagher2015}.

The CIFIST collaboration also compute two further atmospheres for every 3D atmosphere produced with \cobold, for the same $\teff/\logg/\feh$ parameter set as the 3D atmosphere. The first is an averaged 3D or \atd\ model atmosphere. They are computed for every 3D snapshot by spatially averaging the thermal structure of the computational box over surfaces of equal Rosseland optical depth. The other model is an external 1D model atmosphere, computed with the ``Lagrangian hydrodynamical'' (\odx) code. Using model atmospheres produced using the \odx\ code allows for direct evaluation of the 3D effects to a spectrum, as both the 3D and 1D \odx\ atmospheres are calculated over the same finite number opacity bins. Most modern 1D model atmospheres use more sophisticated methods for computing the opacities, which is not currently possible to replicate when working in 3D. Therefore any differences found between the resultant synthesis from the 3D and 1D \odx\ model atmospheres is a direct result of the differences in the underlying physics between the dynamic 3D atmosphere and the static 1D atmosphere, and not due to the constricted opacity physics in the 3D atmospheres.

Most of the atmospheres in the CIFIST grid were originally computed using six opacity bins, but several of the atmospheres in the grid have been recomputed for larger numbers of opacity bins and/or a higher geometrical resolution. If they were available, the atmospheres with larger opacity bin sampling were used to compute the G-band. This is because a lower opacity grid does not accurately model the outermost regions of the star because of the large number of ranges in opacity in the turbulent outer regions of the computational box. The more opacity bins included in the analysis, the better the outer layers are modelled.

\subsection{The G-band line list}
\label{sec:linelist}

We compute a region of the G-band $260\,\AA$ wide between $4140-4400\,\AA$. The line list we constructed to compute the spectral grids was a mixture of atomic and molecular features. We adopted the \element[][12]{CH} information from recently published grids of CH line lists \citep{Masseron2014}, which is publicly available via the molecular line list tool\footnote{\href{http://www.astro.ulb.ac.be/~spectrotools/}{http://www.astro.ulb.ac.be/$\sim$spectrotools/}}. To expedite compute times, we only included \element[][12]{CH} transitions in our line list, providing us with 3395 CH line transitions. The G-band lies very close to the ${\rm H}_\gamma$ balmer line. As such, this transition was included in our line list. We also include several atomic transitions selected from the ``Turn-off Primordial Stars'' (TOPoS) ESO/VLT Large Programme 189.D-0165 \citep{Caffau2013} line lists, which includes a refined atomic line list based on that used in the ``First Stars'' ESO Large Programme 165.N-0276(A) \citep{Bonifacio2009,Cayrel2004}. They were selected based on their impact to a 1D spectrum for a given metallicity such that any atomic transition with an impact greater than 0.1 to the normalised spectrum was considered. For the dwarf model turn-off star atmospheres with $\feh\leq-2.0$, such as those used in the present investigation, it was found that 41 atomic transitions (from 6 species) were needed. For models with $\feh>-2.0$ (and giant, and sub-giant models with $\feh\geq-3.0$) 995 atomic transitions (from 26 species) were required.

\subsection{Setting up Linfor3D}
\label{sec:linfor}

Computing such comprehensive line list over a large wavelength range using classical 1D LTE spectrum synthesis codes is a fairly quick and trivial matter for modern computers. However, the computations undertaken in a 3D spectrum synthesis code require a much larger amount of time (even under LTE as is done here).

Extensive upgrades to Linfor3D means that it now runs under GDL (GNU data language\footnote{\href{http://gnudatalanguage.cvs.sourceforge.net/}{http://gnudatalanguage.cvs.sourceforge.net/}}), eliminating licensing concerns under IDL \citep[see][]{Steffen2014}, meaning that for the first time, large amounts of parallel synthesis could be computed with Linfor3D. The 3D synthesis was split into 10 equally sized wavelength intervals of 26\,\AA. Each wavelength interval's line list also contained the ${\rm H}_{\gamma}$ transition so that the wing of ${\rm H}_\gamma$ was appropriately computed. Additionally, each snapshot was individually computed for every wavelength interval, meaning that up to 200 synthesis were computed for every synthetic G-band we compute. The synthesis was then combined after the computations were complete. This severely reduced the compute time for a single synthetic G-band profile from approximately 100 days to less than 24 hours. The increased production has allowed us to create grids of 3D G-band synthesis for several CNO abundances for many of the CIFIST atmospheres in a relatively short amount of time, giving us an extensive synthetic database to work with.

\subsection{The G-band abundance grid}
\label{sec:abundances}

The models used for the work presented here are listed in Table~\ref{tab:atmgrid}. The chemistry of each model atmosphere input into the spectrum synthesis code was modified for at least six carbon, nitrogen and oxygen abundances, $A({\rm C})$\footnote{$A({\rm X})=\log_{10}{\left(\frac{N({\rm X})}{N({\rm H})}\right)} + 12$. However, the 3D and 1D \odx\ abundances are sometimes referred to separately to distinguish between them, therefore $A({\rm X})$ is sometimes referred to as $A({\rm X})_{\rm 3D}$ or $A({\rm X})_{{\rm 1D},\odx}$ when appropriate.}, $A({\rm N})$ and $A({\rm O})$. The abundance range computed for the ${\rm [Fe/H]} = -2.0$ model atmospheres was set at $6.50\leq A({\rm C})_{\rm 3D}\,[{\rm dex}]\leq 8.50$. This range realistically represents typical abundance ranges for the high plateau CEMP stars \citep{Spite2013}. Spectrum synthesis for the ${\rm [Fe/H]} = -1.0$ and $-3.0$ model atmospheres were computed for the abundance range $6.00\leq A({\rm C})_{\rm 3D}\,[{\rm dex}]\leq 8.00$. The low plateau CEMP stars ($\feh\leq-3.0$) and sub-solar metallicity stars ($-2.0<\feh<0.0$) have typical carbon abundances in this range. All abundance ranges were computed in $\Delta A({\rm C})_{\rm 3D}=0.4\,{\rm dex}$ intervals. The oxygen and nitrogen abundances were scaled with the carbon abundance, relative to the solar CNO abundances. As all models used in this study are metal-poor, the $\alpha$-element abundances, which include oxygen (but not carbon or nitrogen), were enhanced by 0.4\,dex. The C/N\footnote{${\rm X/Y}=N({\rm X})/N({\rm Y})=10^{\left[ A({\rm X})_*-A({\rm Y})_* \right]}$} and C/O ratios were constant at $3.89$ and $0.21$, respectively. Linfor3D uses a chemical network, which is described in \citet[][Sect. 11]{Steffen2014}. All carbon, nitrogen and oxygen molecular species required for the work conducted here are considered by it.

\begin{table}[!ht]
\begin{center}
\caption{List of 3D model atmospheres synthesised for the G-band.}
%\begin{tabular}{@{}p{7em} p{2em} p{1.7em} p{2.2em} p{2.5em} p{4.1em}r@{}}
\begin{tabular}{@{}p{7em} p{1.6em} p{1.5em} p{2.2em} r r@{}}
\hline\hline
Model name & $\teff$ & $\logg$ & $\feh$ & opacity & number of \\
& [K] & [cgs] & [dex] & bins & synthesis \\
\hline
d3t59g40mm10n02 & $5850$ & $4.0$ & $-1.0$ & $6$  & $7$  \\
d3t59g40mm20n02 & $5856$ & $4.0$ & $-2.0$ & $6$  & $7$  \\
d3t59g40mm30n02 & $5846$ & $4.0$ & $-3.0$ & $6$  & $7$  \\
\noalign{\vskip 0.4em}
d3t59g45mm10n01 & $5923$ & $4.5$ & $-1.0$ & $6$  & $7$  \\
d3t59g45mm20n01 & $5923$ & $4.5$ & $-2.0$ & $6$  & $7$  \\
d3t59g45mm30n01 & $5920$ & $4.5$ & $-3.0$ & $6$  & $7$  \\
\noalign{\vskip 0.4em}
d3t63g40mm10n02 & $6236$ & $4.0$ & $-1.0$ & $12$ & $7$  \\
d3t63g40mm20n03 & $6206$ & $4.0$ & $-2.0$ & $12$ & $7$  \\
d3t63g40mm30n02 & $6242$ & $4.0$ & $-3.0$ & $12$ & $12$ \\
\noalign{\vskip 0.4em}
d3t63g45mm10n01 & $6238$ & $4.5$ & $-1.0$ & $6$  & $7$  \\
d3t63g45mm20n01 & $6323$ & $4.5$ & $-2.0$ & $6$  & $7$  \\
d3t63g45mm30n01 & $6273$ & $4.5$ & $-3.0$ & $6$  & $7$  \\
\noalign{\vskip 0.4em}
d3t65g40mm10n01 & $6504$ & $4.0$ & $-1.0$ & $6$  & $7$  \\
d3t65g40mm20n01 & $6534$ & $4.0$ & $-2.0$ & $6$  & $7$  \\
d3t65g40mm30n01 & $6408$ & $4.0$ & $-3.0$ & $6$  & $7$  \\
\noalign{\vskip 0.4em}
d3t65g45mm10n01 & $6456$ & $4.5$ & $-1.0$ & $6$  & $7$  \\
d3t65g45mm20n01 & $6533$ & $4.5$ & $-2.0$ & $6$  & $7$  \\
d3t65g45mm30n01 & $6550$ & $4.5$ & $-3.0$ & $6$  & $7$  \\
\hline
\label{tab:atmgrid}
\end{tabular}
\tablefoot{The model names presented in column 1 represent the current nomenclature preferred by the CIFIST collaboration. They provide a rough model temperature (t), gravity (g), metallicity (m, the extra ``m'' indicates that the metallicity is negative, a ``p'' instead would mean a positive metallicity), and the model version (n). The average temperature for each snapshot selection is listed in column 2. The reason they are not identical for each metallicity and gravity is a direct result of temperature not being a control parameter.}
\end{center}
\end{table}

It was found that varying the nitrogen abundance (i.e. ${\rm C/N}\neq3.89$) had little to no impact on the G-band so no further investigation was warranted for nitrogen. However, it was found that varying the oxygen abundance had a sizeable impact on the G-band. As the oxygen abundance was decreased (increasing C/O), the strength of the CH transitions increased. Therefore one additional synthesis was computed for every atmosphere where the oxygen abundance was scaled down. The carbon and oxygen abundances for the spectral synthesis of the $\feh=-3.0$ and $-1.0$ models were $A{\rm (C)} = 6.80$ and $A{\rm (O)} = 6.20$, while the abundances for the spectral synthesis of the $\feh=-2.0$ models were $A{\rm (C)} = 7.30$, $A{\rm (O)} = 6.70$. Therefore, the C/O ratio increased from $0.21$ to $3.98$. The nitrogen abundances were still scaled with the carbon abundances.

To further test the effect that the oxygen abundance has on the 3D synthesis, we computed additional synthesis using a single model atmosphere, d3t63g40mm30n02. In this case the oxygen abundance was fixed at 6.20\,dex, while the carbon abundance was synthesised for the same range as before, $6.00\leq A({\rm C})_{\rm 3D}\,[{\rm dex}]\leq 8.00$, with $\Delta A({\rm C})_{\rm 3D}=0.4\,{\rm dex}$, such that $0.63\leq {\rm C/O}\, \leq63.10$. This increased the number of synthesis for this atmosphere from six to 12. The reasons for this become apparent in Sect.~\ref{sec:oxygen}.

This paper represents a purely theoretical exercise, and as such we describe how the 3D synthesis behaves relative to the 1D \odx\ synthesis. This required the creation of a grid of 1D \odx\ synthetic spectra. We stress that the 1D \odx\ model atmospheres represent a preferred choice when comparing to a 3D model atmosphere over others, such as ATLAS \citep{Kurucz2005} or MARCS \citep{Gustafsson2008} for the reasons described above. As such, they are used to produce all the 1D synthesis used in this work. For every atmosphere listed in Table~\ref{tab:atmgrid} a grid of 51 1D LTE synthetic spectra were produced by Linfor3D. Each grid spanned a large carbon abundance range, $5.50\leq \acod\,[{\rm dex}]\leq10.50$ with $\Delta\acod = 0.10$\,dex, which would cover the abundance range computed for the 3D grid. While the oxygen abundance is important to the G-band in 3D, the oxygen abundance has almost no effect to the G-band under 1D. Therefore the oxygen (and nitrogen) abundances are inconsequential in 1D and are not reported further. However, for reference, we scaled the oxygen and nitrogen abundances with the carbon abundance, maintaining constant C/O and C/N ratios ($0.21$ and $3.89$, respectively). It should be noted that \citet{Collet2007} found similar effects to individual CH and OH lines by changing the C/O ratio in giants as we find for the G-band in dwarfs stars presented here.

\begin{figure*}[!htp]
	\begin{center}
	\includegraphics[scale=0.515]{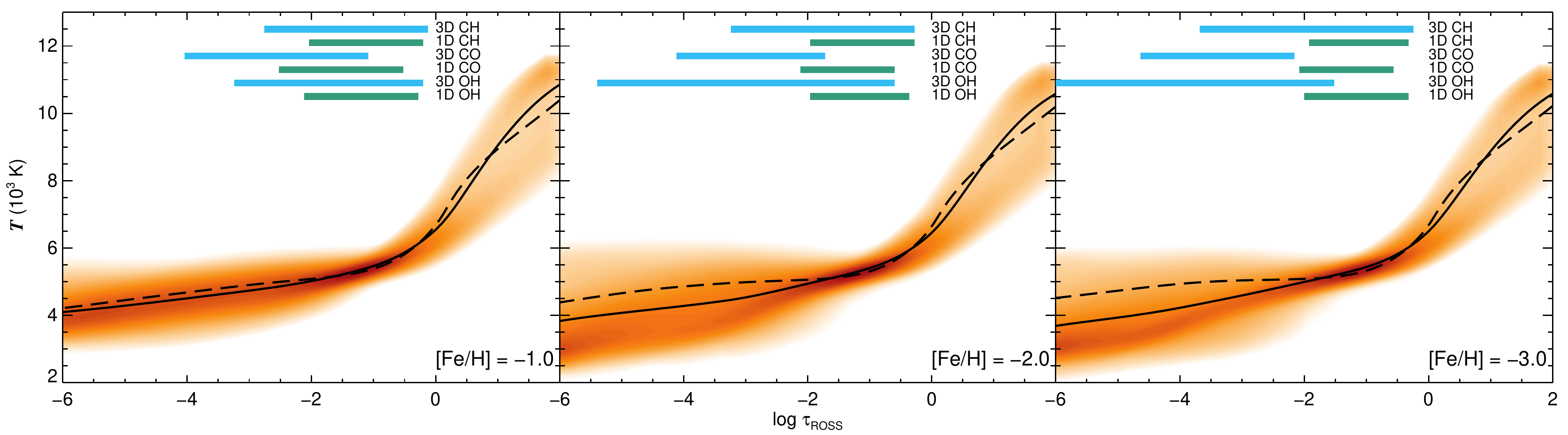}
	\caption{The temperature structures of the 3D \cobold\ (orange density map), \atd\ (solid black line) and 1D \odx\ (dashed black line) for the d3t63g40mm10n02, d3t63g40mm20n03 and d3t63g40mm30n02 model atmospheres. Darker regions of the density map indicate more frequently sampled temperatures. The 1D \odx\ model atmospheres are computed assuming $\alpha_{\rm MLT}=1.0$. Contribution functions of a low-excitation CH, CO and OH line computed in 3D (light blue) and 1D (green) are included to demonstrate the differences between the 3D and 1D \odx\ forming regions and how the metallicity impacts the depth of formation. Their CNO abundances are identical for all metallicities; $\actd=\acod=6.80$, $A({\rm N})_{\rm 3D}=A({\rm N})_{{\rm 1D},\odx}=6.21$, $\aotd=\aood=7.47$.}
	\label{fig:tempstruct}
	\end{center}
\end{figure*}

\section{Formation of the G-band}
\label{sec:gbandformation}

The majority of the G-band consists of CH molecule transitions. There are also a limited number of atomic species transitions in the wavelength range computed for this work. \citet{Wedemeyer2005} detail the implementation of time-dependent chemical reaction networks in \cobold. They also examine CO in a two-dimensional (2D) atmosphere, created from slices of the 3D \cobold\ atmosphere. Their analysis indicated that the population of CO is strongest in the cooler regions of their models and also demonstrate that CO is extremely important in the formation of carbon and oxygen molecules (see their Fig.~1 and Table~2).

Consequently, it is important to examine the interrelationships between the carbon and oxygen based molecules in the 3D atmosphere so that the behaviours observed in the G-band can be fully understood. Therefore in this section we examine how temperature structures in the 3D and 1D \odx\ model atmospheres influence the resultant molecular number densities and contribution functions of three molecules, CH, CO and OH, exploring the differences between them, as well as explore how varying the C/O ratio impacts these properties.

\subsection{Temperature structure effects}
\label{sec:tempstructure}

Previous works have shown that metal-poor 3D model atmospheres are, in general, cooler in the outer atmosphere than the equivalent 1D atmosphere. Fig.~\ref{fig:tempstruct} depicts the 3D, \atd\ and equivalent 1D \odx\ temperature structures for the model atmospheres d3t63g40mm10n02, d3t63g40mm20n03 and d3t63g40mm30n02. These model atmospheres share the same gravity ($\logg=4.0$), and very similar temperatures, but have different metallicities. It is shown that the temperature differences between the 1D \odx\ and \atd\ atmospheres (and hence the 3D atmospheres) increase as the metallicity is decreased at low $\logtr$. In the $\feh=-3.0$ atmosphere it is shown that the 1D \odx\ atmosphere is hotter than the \atd\ and the densest regions of the 3D atmospheres (represented by the darker shaded areas) by $\sim100-840$\,K over the region $-6\leq\logtr\leq-2$. Over the same region, the temperatures between the 1D \odx\ and 3D $\feh=-1.0$ model atmospheres differ by $\sim65-100$\,K, and the 1D \odx\ (and \atd) atmosphere (almost) trace densest regions of the 3D temperature structure. The cooling effect seen in the 3D $\feh=-3.0$ and $-2.0$ model atmospheres can be attributed to inefficient radiative heating of the outermost layers of low metallicity model atmospheres relative to hydrodynamic cooling via overshooting \citep{Asplund1999,Asplund2001}. As the metallicity is increased from $1/100^{\rm th}$ to $1/10^{\rm th}$ solar, the radiative heating becomes more efficient, heating up the outer layers of the 3D atmosphere. Therefore, the outer layers of all three temperature structures are similar.

Further temperature deviations between the 1D \odx\ and 3D model atmospheres are seen in the deeper layers of the star. This can be controlled somewhat by the mixing length parameter, $\alpha_{\rm MLT}$, in the 1D \odx\ model atmosphere. When $\alpha_{\rm MLT}$ is altered, the temperature structures change dramatically \citep[see][Fig.~2]{Gallagher2015}. For lines that form in deep layers of the star ($\logtr\gtrsim0$), the differences in temperature will impact their formation. However, molecules do not form in the deeper regions of a star due to molecular dissociation. To illustrate this point, Fig.~\ref{fig:tempstruct} also presents the formation regions for a low excitation CO, CH and OH line in both 3D and 1D. As low excitation lines are known to form higher up in the 3D atmosphere, we examined higher excitation lines in both the 1D \odx\ and 3D model atmospheres and found that no lines indicated any significant formation in regions deeper than $\logtr = 0$. As such, line formation would not show any notable sensitivity to the mixing length parameter, so this is not considered further. We set $\alpha_{\rm MLT}=1.0$ for every 1D \odx\ atmosphere used in this study.

The forming regions of the three molecular lines synthesised in 1D and 3D in Fig.~\ref{fig:tempstruct} are determined by their equivalent width contribution functions \citep[see][for the formal definition]{Steffen2014}. All of these lines have low excitation potentials, $0.01\leq\chi\,({\rm eV})\leq0.19$. It was pointed out in \citet{Gallagher2015} that the contribution functions calculated from low excitation atomic lines in 3D are very different to the counterpart 1D contribution functions in metal-poor atmospheres. We see the same effect in the low excitation carbon and oxygen molecules too (right panel, Fig.~\ref{fig:tempstruct}). \citet{Dobrovolskas2013} show that as the metallicity is increased in giant model atmospheres, low excitation \ion{Fe}{i} lines begin to form over the same regions in the 3D atmosphere as they do in the equivalent 1D atmosphere, whereas in the metal-poor atmosphere the low excitation \ion{Fe}{i} lines form further out in 3D. We see a similar effect in the CO, CH and OH molecules in the dwarf models in Fig.~\ref{fig:tempstruct}; the higher metallicity model atmospheres show very similar 3D and 1D line formation regions, which diminishes as the model atmospheres decrease in metallicity. The apparent differences in line formation between the high and low metallicity model atmospheres will have a significant impact on any 3D abundance corrections reported below.

\subsection{Molecular number densities}
\label{sec:levpops}

\begin{figure*}[!th]
	\includegraphics[scale=1]{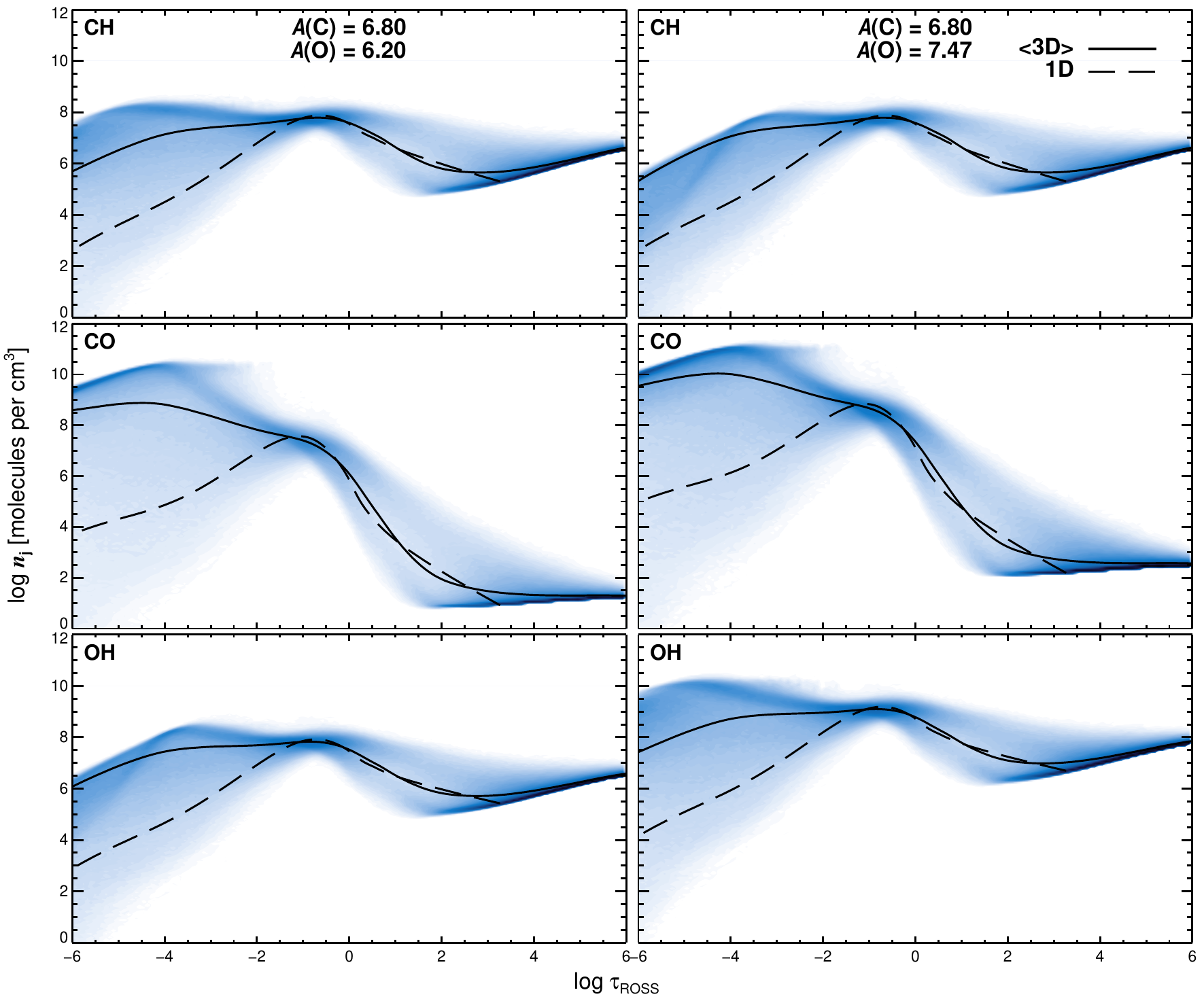}
	\caption{molecular number densities of CH, CO and OH in the d3t63g40mm30n02 atmosphere (blue density map) for fixed $A({\rm C})$ and two values of C/O. The left panels present the molecular number densities for an atmosphere with ${\rm C/O}=3.98$ and the right panels an atmosphere with ${\rm C/O}=0.21$. The \atd\ (solid black line) and 1D \odx\ (dashed black line) molecular number densities are also included. For clarity, each panel has the same axis.}
	\label{fig:levpop}
\end{figure*}

The difference in temperatures between the metal-poor 1D \odx\ and 3D atmospheres have large effects on the molecular number densities of the carbon and oxygen molecules, which in turn affect the G-band. Fig.~\ref{fig:levpop} shows how the change in $A({\rm O})$ affects the molecular number densities of CH, CO and OH in the d3t63g40mm30n02 model atmosphere. (Figs.~\ref{afig:levpop1}~\&~\ref{afig:levpop2} present the same plots as Fig.~\ref{fig:levpop} for the $\feh=-1.0$ \& $-2.0$ model atmospheres, respectively.) The panels on the left present a case where the oxygen abundance is depleted relative to carbon, such that ${\rm C/O}=3.98$. The panels on the right present a metal-poor atmosphere with a scaled solar abundance pattern, ${\rm C/O}=0.21$. It is clear from the figure that CO molecules dominate in this metal-poor atmosphere in both abundance regimes. It is also evident that altering the oxygen abundance in the 3D atmospheres not only directly impacts the CO and OH populations, but also has some secondary impact on the CH molecular number densities as well. However, this effect is not seen in the \atd\ and 1D CH molecular number densities.

\subsubsection{CO molecules}
\label{sec:COmol}

The CO molecule is one of the most durable in nature, owed to the triple bond (two covalent bonds and one dative) between the carbon and oxygen atoms. This gives it a dissociation energy of 11.092\,eV, far larger than CH (3.465\,eV) and OH (4.392\,eV) \citep{Tatum1966}. As such, CO molecule formation dominates over all other associated carbon, oxygen and nitrogen molecules as the numbers of CO molecules in this metal-poor 3D stellar atmosphere are $\sim2$ orders of magnitude larger in the typical line forming regions ($-6\lesssim\logtr\lesssim0$) than the numbers of CH or OH molecules. Despite the outer atmosphere having a low density, the temperatures in the 3D atmosphere are low enough so that molecular formation can occur. Therefore CO plays a considerable role in the formation of carbon-bearing molecules in this metal-poor atmosphere, which includes CH, making oxygen abundances important to the formation of CH. This is not seen in classical 1D model atmospheres (and hence the resultant 1D spectrum synthesis), a reason owed most likely to the higher temperatures in the 1D model coupled with the lower densities found in the outer atmosphere which hinders CO formation, forcing the majority of CO to form deeper in the 1D atmosphere where densities are higher. Therefore we see a monotonic increase in the number of CO molecules in the range $-4\lesssim\logtr\lesssim2$ in the 3D atmosphere. In the deepest layers ($-1\lesssim\logtr\leq2$) of the 1D \odx\ atmosphere, the CO molecular number densities trace the 3D, but in the outer layers of the atmosphere ($\logtr>-1$) they dramatically decrease. This means that CO transitions have a much larger range to form over in the 3D atmosphere, leading to stronger lines.

Decreasing the C/O ratio from $3.98$ to $0.21$ (right and left panels of Fig.~\ref{fig:levpop} respectively) increases CO molecular number densities by approximately one order of magnitude in all three types of atmosphere. The pattern of behaviour of the CO molecule populations remain unaffected, however. In the higher metallicity model depicted in Fig.~\ref{afig:levpop1}, the CH molecular number densities are largely  independent of the C/O ratio. This means that in the more metal-rich atmospheres the oxygen abundance has almost no impact on the G-band. When we compare Fig.~\ref{fig:levpop} with Fig.~\ref{afig:levpop1} we see that the metal-poor model at $\feh=-3.0$, which has cooler upper photospheric layers, allows for more CO formation regions in the outer atmosphere, leading to a cumulative increase in the amount of CO absorbers, whereas the hotter $\feh=-1.0$ atmosphere begins to disassociate the CO molecules in these regions meaning that any increase to the number of CO molecules is owed to the temperature and density fluctuations in the 3D atmosphere.

\subsubsection{OH molecules}
\label{sec:OHmol}

The OH molecules also play a role in CH formation. Like CO, the increase or decrease of the oxygen abundance increases or decreases the number of OH molecules in all three types of atmosphere, which is expected. The two bottom panels of Fig.~\ref{fig:levpop} depict the effect of altering the oxygen abundance by $\sim\pm1.3$\,dex, while fixing the carbon abundance. Hence the C/O ratio has been shifted $\mp3.77$. Increasing the oxygen abundance has an effect on OH in the 1D and \atd\ models; both have shown a general increase in OH molecular density. This is to be expected as changing the oxygen abundance directly influences the molecular number densities of OH. The increase in the 3D molecular number densities behaves in much the same way in the deeper regions of the 3D atmosphere as it does in the 1D and \atd\ atmospheres. As $\logtr$ falls below $-3.0$, the 3D molecular number densities of OH further increase relative to the 1D model and even the \atd\ model. This will be reflected in the 3D synthesis by increased line formation in the outer atmosphere when the C/O ratio is low. In the $\feh = -2.0$ model (Fig.~\ref{afig:levpop2}), it is found that the increase in oxygen abundance has the same overall effect to the 3D and 1D OH molecular densities as is described in the $\feh=-3.0$ model. It does not have the same effect in the \atd\ model, however. This is due to the fact that the \atd\ model is generally hotter than the 3D temperature structure (Fig.~\ref{fig:tempstruct}). Therefore \atd\ OH molecular formation is reduced, relative to that found in the $\feh=-3.0$ model. When the $\feh=-1.0$ model is examined in the same way, we find that OH molecular formation in the 3D behaves like the 1D and \atd\ molecular formation such that no further increase in population density is found.

\subsubsection{CH molecules}
\label{sec:CHmol}

The CH molecules are responsible for the G-band. Understanding their behaviour as the CNO abundances vary is of primary concern in this work. We have just seen that there is a secondary effect to CH molecule formation due to the oxygen abundance, which reduces the number of carbon atoms available to form CH because of CO, as well as the primary effect of altering the carbon abundance, but only under full 3D. This is clear when the top two panels of Fig.~\ref{fig:levpop} are studied in detail; the 1D \odx\ and \atd\ CH populations remain almost identical to one-another when $A({\rm O})$ is changed. However, there is a dramatic increase to the CH formation in the outermost part of the 3D atmosphere ($\logtr\leq-3.0$) as $A({\rm O})$ is reduced. The behaviour of CH in $\feh = -2.0$ model (Fig.~\ref{afig:levpop2}) is fairly similar to that described in the $\feh = -3.0$ model. However, the 3D CH molecular population in the $\feh=-1.0$ model shows very little influence by the change in the oxygen abundance. The hotter temperatures in this model restricts the production of CO, reflected by the large difference in CO number densities between the $\feh=-1.0$ and $-3,0$ models, such that CH is less sensitive to the oxygen abundance because of the increase in free carbon atoms.

\subsubsection{Combined effects}
\label{sec:allmol}

It is clear that the oxygen abundance (and hence the C/O ratio) is crucial to the reported carbon abundance found from analysis of the G-band under 3D. Increasing the oxygen abundance serves to increase the number of oxygen atoms available to form CO and OH. Additionally, the increasing OH population removes oxygen atoms in CO formation, but this effect is expected to be small, given the preference to form CO over OH in a stellar atmosphere, as judged by the large differences in CO molecules relative to OH, at any oxygen abundance. The CO molecules remove carbon atoms for CH formation resulting in a decreased amount of CH formation as oxygen is increased at any given carbon abundance. The response observed in the CH and OH molecules as the oxygen abundance is varied suggests that they are both highly dependent on the formation of CO.

This effect decreases as metallicities are increased. When $\feh=-1.0$, the CO populations are smaller, relative to those in the $\feh=-3.0$ model. While the effect of changing the oxygen abundance does not change the behaviour of CO molecule formation, the overall reduction in the numbers of CO molecules, relative to the carbon and oxygen abundance means that CH and OH becomes less sensitive to CO as there are greater numbers of carbon and oxygen atoms to form CH and OH, respectively. Therefore, the C/O ratio becomes less important to CH and OH.

\subsection{Line formation}
\label{sec:formationdepth}

\begin{figure*}[!th]
	\includegraphics[scale=1]{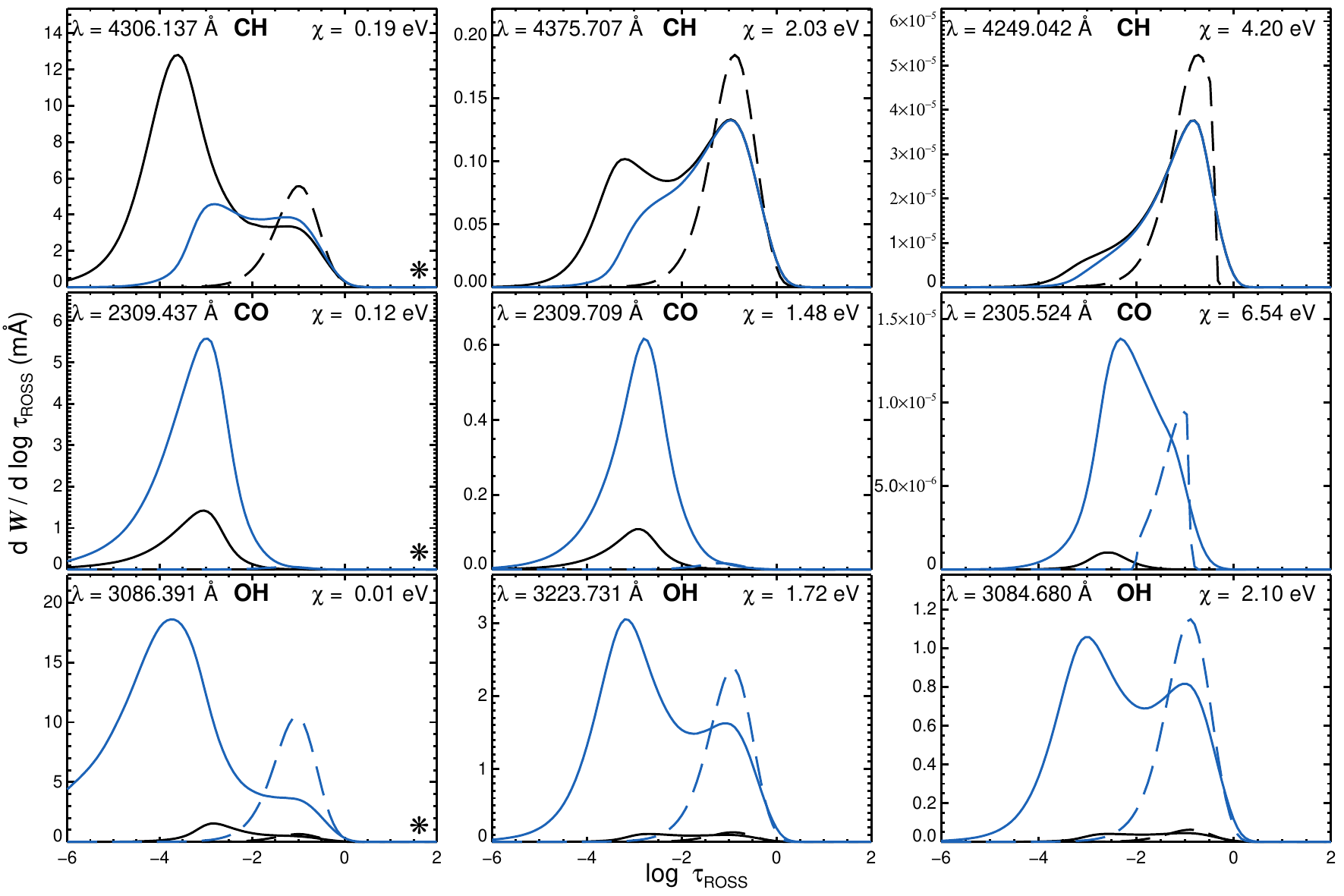}
	\caption{CH, CO and OH equivalent width contribution functions in 3D (solid lines) and and 1D (dashed lines) calculated for lines with different excitation potentials in the d3t63g40mm30n02 atmosphere. $\actd=\acod= 6.80\,{\rm dex}$ with $\aotd=\aood=6.20\,{\rm dex}$ (black) and $7.47\,{\rm dex}$ (blue). The oxygen abundance does not affect 1D CH contribution function. Those panels marked with a star are also depicted as horizontal bars in Fig.~\ref{fig:tempstruct}.}
	\label{fig:contfunc}
\end{figure*}

The 3D temperature structure seen in Fig.~\ref{fig:tempstruct} shows large horizontal fluctuations in temperature. It was shown that these temperature fluctuations lead to additional cool regions within the atmosphere where molecules can form, and hence can contribute to the overall line strength. In Fig.~\ref{fig:contfunc}, we present the equivalent width contribution functions for three CO, CH and OH lines forming in the d3t63g40mm30n02 model atmosphere. (Figs.~\ref{afig:contfunc1}~\&~\ref{afig:contfunc2} depict the associated equivalent width contribution functions for the $\feh=-1.0$ and $-2.0$ model atmospheres, respectively, for comparison.) The excitation potentials of the three selected lines in each case serve to illustrate the effect it has to line formation. The blue and black contribution functions demonstrate the effect that the C/O ratio has on feature formation by fixing the carbon abundance, $A({\rm C}) = 6.80\,{\rm dex}$, and varying the oxygen abundance, $A({\rm O}) = 7.47\,{\rm dex}$ (blue) and $A({\rm O}) = 6.20\,{\rm dex}$ (black). The dashed lines in each plot represent the equivalent synthesis from the 1D \odx\ atmosphere with the same abundances. The changes in the oxygen abundance do not have the same impact on CH features in the more simplistic 1D \odx\ synthesis that we see in the 3D synthesis, meaning that their equivalent widths remain the same. Therefore, we have only plotted a single example to make the figure as clear as possible.

When we examine Fig.~\ref{fig:contfunc} in detail, we see evidence of the extra formation regions (described in Sect.~\ref{sec:levpops}) contributing to the overall line strength because of the extended cool regions where the lines form in 3D, relative to the 1D. This also explains why the overall line strength in the 3D synthesis (the integral of the contribution function over $\logtr$ in this case) is much larger in the 3D than the 1D for a given CNO abundance and why one must reduce said CNO abundances in 3D to reproduce the same line strength as the 1D.

\subsubsection{CO line formation}
\label{sec:COform}

CO lines are extremely weak. This is owed to low transition probabilities. In 3D, it is generally expected that CO lines become stronger, due to the effects presented in Sect.~\ref{sec:levpops}. However, it is difficult to find real CO lines that are strong enough to examine the contribution functions in detail. Therefore, we have artificially enhanced the $\loggf$ values expected in theory by two orders of magnitude in both the 3D and 1D synthesis. This increases the transition probabilities of the CO lines presented, increasing the overall strength of their contribution at a given depth in the atmosphere. However, the strength of these lines will not be consistent with what is observed in nature. As our explanations are qualitative, this does not present us with any difficulties. Nevertheless, formation of CO molecules in 3D atmospheres does indeed appear to lead to stronger line profiles (up to $\sim600$ times larger).

From Fig.~\ref{fig:contfunc} it is clear that CO molecules have some sensitivity to their excitation potential, shifting the majority of the transition formation towards the outer atmosphere as $\chi$ decreases. This effect is seen for all three molecules in 3D, but its impact is smallest in the CO transitions. A similar effect was reported in \citep{Gallagher2015} in iron line formation, which they suggest is most likely a result of assuming LTE line formation in regions of the star where non-local temperature effects are important. For the moment, computing \emph{full} 3D non-local thermodynamic equilibrium (NLTE) departures for molecular data (or indeed iron line data) has not been attempted in great detail because of the complexities involved in its execution. However, considerable progress has been made on 3D-NLTE iron line formation, the details of which can be found in \citet{Holzreuter2013,Holzreuter2015}.

When the oxygen abundance is increased in 3D, the strength of the CO transitions increases dramatically. This is expected because the molecular number density of CO increases as the oxygen abundance is increased. This effect is the same in 1D (dashed lines), however, the overall strength of the lines is so weak that they are not visible in all but one panel in Fig.~\ref{fig:contfunc}. As the metallicity is increased (Figs.~\ref{afig:contfunc1} \&~\ref{afig:contfunc2}) it is clear that the formation occurs deeper in the 3D atmosphere, until the 1D and 3D formation depths are similar. However, the effect of the 3D temperature fluctuations result in much stronger CO lines compared with the equivalent 1D formation.

\subsubsection{OH line formation}
\label{sec:OHform}

Unsurprisingly, the OH transitions reduce dramatically when the oxygen abundance is reduced. The lines we use to illustrate OH contributions throughout the stellar atmosphere have a limited excitation potential range (0.010, 1.723 and 2.095 eV from left to right, respectively). Though they represent some of the largest differences found for OH transition excitation potentials within the UV, visible and near-IR. However, it is enough to illustrate variation in formation depth. While the 1D lines form over the typical range $-3\lesssim\logtr\lesssim0$, the 3D profiles form further out in the atmosphere as the excitation potentials are reduced, resulting in part of the formation of the line with the lowest excitation potential occurring outside the computational box making the abundances attained from such a line unreliable in 3D.

Another notable effect demonstrated by the 3D OH lines is the gradual increase in lines strength (with decreasing $\chi$), relative to the 1D lines. This has been seen and noted in iron lines presented in \citet{Gallagher2015}, of which they report a large excitation potential dependency on abundance. This will most likely be seen in the OH lines as well. Finally, and like the CO transitions just discussed, increasing the metallicity pushes the formation of the OH transitions to regions where the 1D formation takes place (Figs.~\ref{afig:contfunc1} \&~\ref{afig:contfunc2}).

\subsubsection{CH line formation}
\label{sec:CHform}

It is clear that the depth of formation of CH lines follow their excitation potential; high excitation lines form deep in the atmosphere, low excitation lines form mostly in the outer atmosphere where the temperatures are cool. It is also clear that oxygen abundances have a variable impact on CH formation, as judged from the large differences between the blue (largest $A({\rm O})$) and black (smallest $A({\rm O})$) lines in the high and low excitation lines.

The low excitation lines form mostly in the outer atmosphere. This is because the lower level of these transitions is well populated in the outer atmosphere. Therefore such levels are fully populated and require very little contribution in the deeper atmosphere. These lines show a high sensitivity to the change in oxygen abundance, and hence the C/O ratio. The lower temperatures and densities in the outer regions of the 3D atmosphere allow for large temperature and density fluctuations. As the C/O ratio is decreased (i.e. the oxygen abundance is increased) greater numbers of oxygen atoms form larger quantities of CO in the cooler regions of the atmosphere at the expense of CH.

The high excitation lines are populated by higher energy transitions, which are found in the deeper regions of the atmosphere, where temperatures and densities are higher. Temperature fluctuations in these regions of the 3D model atmospheres ($0\leq\logtr\leq-2$) are small, relative to the outer layers of the atmosphere. As such, 3D CH populations in these horizontal layers are similar to those found in the \atd\ and 1D \odx\ atmospheres (Fig.~\ref{fig:levpop}), meaning that 3D CH formation is restricted, relative to formation occurring in the outer layers of the 3D model. Additionally, the sensitivities to the C/O ratio are smaller in the deeper regions of the atmosphere, where these high excitation lines form (see Appendix~\ref{appdx:COratio} for a simple example). Therefore, the high excitation CH lines have smaller sensitivity to oxygen abundances, which is shown in the top right panel of Fig~\ref{fig:contfunc}. This means that the C/O ratio sensitivity shown by CH is related to the temperature and density fluctuations in the 3D atmosphere.

When the higher metallicity figures are examined, the CH transitions shift to deeper layers in the atmosphere, as was seen in the CO and OH lines.

\subsection{Effect on 3D line formation vs 1D line formation}

For all three species of molecule, the same effect is seen under 3D. As the excitation potential reduces, lines forming in the 3D atmosphere form towards the outer atmosphere and their strength increases, relative to the equivalent line forming in the 1D atmosphere. The oxygen abundance is of primary importance to CO and OH line formation in both 1D and 3D, but in general the lines are stronger in 3D for a given C and O abundance. Increasing the oxygen abundance pushes the line formation of the OH molecules towards the outer regions of the 3D atmosphere, particularly in cases where the excitation potential is small due to the reasons discussed above and also due in part to line saturation, but it does not have the same impact on the CO molecules. The formation of CH features occur in deeper layers of the atmosphere in instances where the oxygen abundance is larger. However, when the oxygen abundance is depleted, the formation of CH features is pushed towards the outer regions of the 3D atmosphere, while the OH features begin to form deeper in the atmosphere. This demonstrates an anti-correlated behaviour, which is discussed further in Sect.~\ref{sec:discussion} and Appendix~\ref{appdx:COratio}. Neither effect is seen in the 1D case as it is clear from Fig.~\ref{fig:levpop} that CH, CO and OH molecular number densities are small in the outer layers of the atmosphere. Comparisons between the \atd\ and 1D \odx\ temperature structures would suggest that the primary reason for this is the increased temperatures in the outer layers of the 1D atmosphere.

\section{Spectral synthesis analysis}
\label{sec:gband_analysis}

In this section, we present 3D corrections attained from fitting the 3D LTE G-band spectra with synthesis computed using the 1D \odx\ model sharing the same parameters (temperature, gravity, metallicity) and the chemical composition, except for the carbon abundance. For this, we treated the 3D spectra heuristically as observational spectra of infinite signal-to-noise. To obtain the $\chi^2$ minimisation, the $\chi^2$ code allowed small offsets in carbon abundance, the macroturbulence and wavelength shift, as it is expected that the dynamic nature of the 3D atmospheres will lead to small convective wavelength shifts. The microturbulence of the 1D synthesis was kept fixed. The fits were done for the six 3D spectra where we have scaled the nitrogen and oxygen abundances with the carbon abundance, and hence have a fixed C/O ratio (Sect.~\ref{sec:3dcorrection}). This is useful if one does not know the oxygen abundance of the star (which is frequently the case in metal-poor stars). The 1D spectra were also fit to the 3D spectrum where the oxygen abundance has been depleted, or the C/O ratio has been enhanced (Sect.~\ref{sec:oxygen}). This presents us with the impact the C/O ratio has on the 3D correction, and hence allows one to scale the 3D corrections presented in Sect.~\ref{sec:3dcorrection} if the oxygen abundance is known.

\subsection{3D abundance corrections}
\label{sec:3dcorrection}

\begin{figure*}[ht]
	\includegraphics[scale=1]{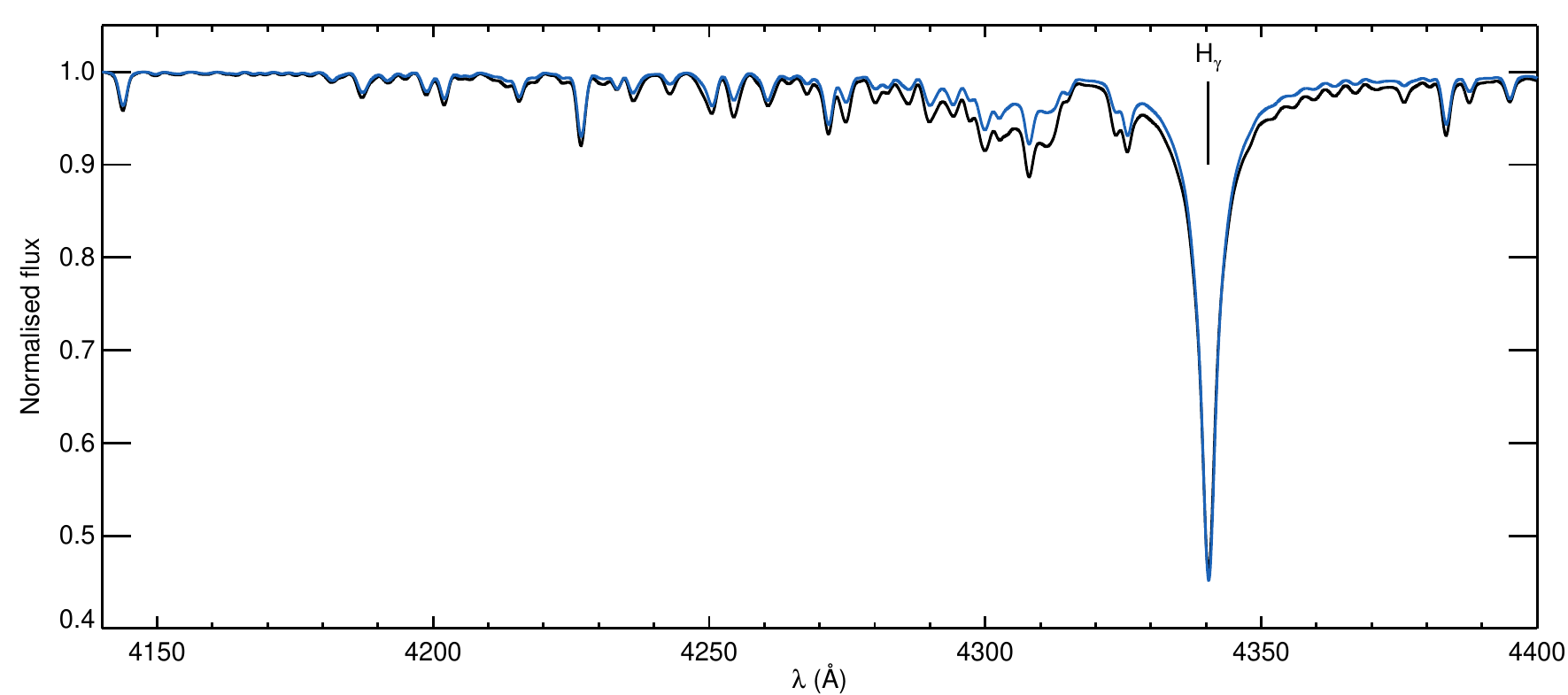}
	\caption{The G-band for a 3D synthesis (black line) and a 1D synthesis (blue line). Both synthetic bands have the same CNO abundances, $A({\rm C})_{\rm 3D}=A({\rm C})_{{\rm 1D},\odx}=6.80$, $A({\rm N})_{\rm 3D}=A({\rm N})_{\rm 1D,\odx}=6.21$, and $A({\rm O})_{\rm 3D}=A({\rm O})_{{\rm 1D},\odx}=7.47$.}
	\label{fig:gband_1d3d}
\end{figure*}

From the work presented in Sect.~\ref{sec:gbandformation}, it is expected that the 3D and 1D line profiles have differing line strengths for a given CNO abundance. Fig.~\ref{fig:gband_1d3d} presents the G-band synthesis for the d3t63g40mm30n02 model atmosphere and demonstrates that for a given CNO abundance (we stress that the nitrogen abundance is inconsequential to the G-band), the 3D profile is much stronger than the 1D profile. So that it is clear where the impact is largest to the G-band, we have set the instrumental broadening of the synthesis so that it is comparable to SDSS spectra ($160\,\kms$). It is also expected that the turbulent motions in the 3D atmosphere will result in a small wavelength shift. The black line represents the 3D line profile and the blue line represents the 1D. It is immediately obvious that the band head ($4280-4310\,\AA$) is particularly affected by the 3D effects. Given this, we expect an abundance correction to the 1D synthesis to reproduce the 3D synthesis, and based on evidence presented in Sect.~\ref{sec:gbandformation} we expect the resulting abundance corrections will increase as the metallicity decreases.

\begin{table*}[ht]
\begin{center}
\caption{Abundance corrections, $\ctd=\actd-\acod$, for the 3D model atmospheres listed in Table~\ref{tab:atmgrid}.}
\begin{tabular}{l c c c c c c c c c r}
\hline\hline
$A({\rm C})_{\rm 3D}$ & $A({\rm N})_{\rm 3D}$ & $A({\rm O})_{\rm 3D}$ & \multicolumn{8}{c}{$\ctd$} \\
\hline
&&&\multicolumn{2}{c}{$\teff\approx5900$\,K} && \multicolumn{2}{c}{$\teff\approx6300$\,K} && \multicolumn{2}{c}{$\teff\approx6500$\,K} \\
&&& $\logg=4.0$ & $\logg=4.5$ && $\logg=4.0$ & $\logg=4.5$ && $\logg=4.0$ & $\logg=4.5$ \\
\cline{4-5} \cline{7-8} \cline{10-11} \vspace{-0.8em}\\
\hline
\multicolumn{11}{c}{$\feh=-1.0$}\\
\hline
$6.00$ & $5.41$ & $6.67$ & $-0.10$ & $-0.16$ && $-0.18$ & $-0.15$ && $-0.30$ & $-0.36$ \\
$6.40$ & $5.81$ & $7.07$ & $-0.04$ & $-0.10$ && $-0.11$ & $-0.10$ && $-0.20$ & $-0.22$ \\
$6.80$ & $6.21$ & $7.47$ & $-0.01$ & $-0.05$ && $-0.06$ & $-0.09$ && $-0.10$ & $-0.13$ \\
$7.20$ & $6.61$ & $7.87$ & $+0.04$ & $-0.01$ && $-0.05$ & $-0.03$ && $-0.09$ & $-0.10$ \\
$7.60$ & $7.01$ & $8.27$ & $+0.08$ & $+0.05$ && $-0.02$ & $-0.02$ && $-0.07$ & $-0.07$ \\
$8.00$ & $7.41$ & $8.67$ & $+0.12$ & $+0.11$ && $+0.00$ & $+0.03$ && $-0.04$ & $-0.02$ \\
\hline
\multicolumn{11}{c}{$\feh=-2.0$}\\
\hline
$6.50$ & $5.91$ & $7.17$ & $-0.17$ & $-0.17$ && $-0.28$ & $-0.32$ && $-0.24$ & $-0.36$ \\
$6.90$ & $6.31$ & $7.57$ & $-0.11$ & $-0.07$ && $-0.24$ & $-0.25$ && $-0.21$ & $-0.30$ \\
$7.30$ & $6.71$ & $7.97$ & $-0.01$ & $+0.00$ && $-0.19$ & $-0.18$ && $-0.17$ & $-0.24$ \\
$7.70$ & $7.11$ & $8.37$ & $+0.08$ & $+0.13$ && $-0.12$ & $-0.08$ && $-0.13$ & $-0.17$ \\
$8.10$ & $7.51$ & $8.77$ & $+0.18$ & $+0.22$ && $-0.05$ & $+0.00$ && $-0.07$ & $-0.09$ \\
$8.50$ & $7.91$ & $9.17$ & $+0.28$ & $+0.35$ && $+0.02$ & $+0.02$ && $+0.02$ & $-0.01$ \\
\hline
\multicolumn{11}{c}{$\feh=-3.0$}\\
\hline
$6.00$ & $5.41$ & $6.67$ & $-0.31$ & $-0.28$ && $-0.37$ & $-0.45$ && $-0.39$ & $-0.49$ \\
$6.40$ & $5.81$ & $7.07$ & $-0.20$ & $-0.19$ && $-0.31$ & $-0.37$ && $-0.34$ & $-0.43$ \\
$6.80$ & $6.21$ & $7.47$ & $-0.14$ & $-0.09$ && $-0.25$ & $-0.29$ && $-0.29$ & $-0.35$ \\
$7.20$ & $6.61$ & $7.87$ & $-0.03$ & $+0.02$ && $-0.20$ & $-0.20$ && $-0.22$ & $-0.28$ \\
$7.60$ & $7.01$ & $8.27$ & $+0.06$ & $+0.13$ && $-0.14$ & $-0.10$ && $-0.17$ & $-0.22$ \\
$8.00$ & $7.41$ & $8.67$ & $+0.15$ & $+0.23$ && $-0.08$ & $-0.02$ && $-0.11$ & $-0.13$ \\
\hline
\label{tab:3Dcorrections}
\end{tabular}
\tablefoot{The C/N and C/O ratios are constant at 3.89 and 0.21, respectively. The error assigned to every correction listed is $\pm0.05$, for the reasons explained in the text.}
\end{center}
\end{table*}

\begin{figure*}[!th]
\begin{center}
	\includegraphics[scale=0.71]{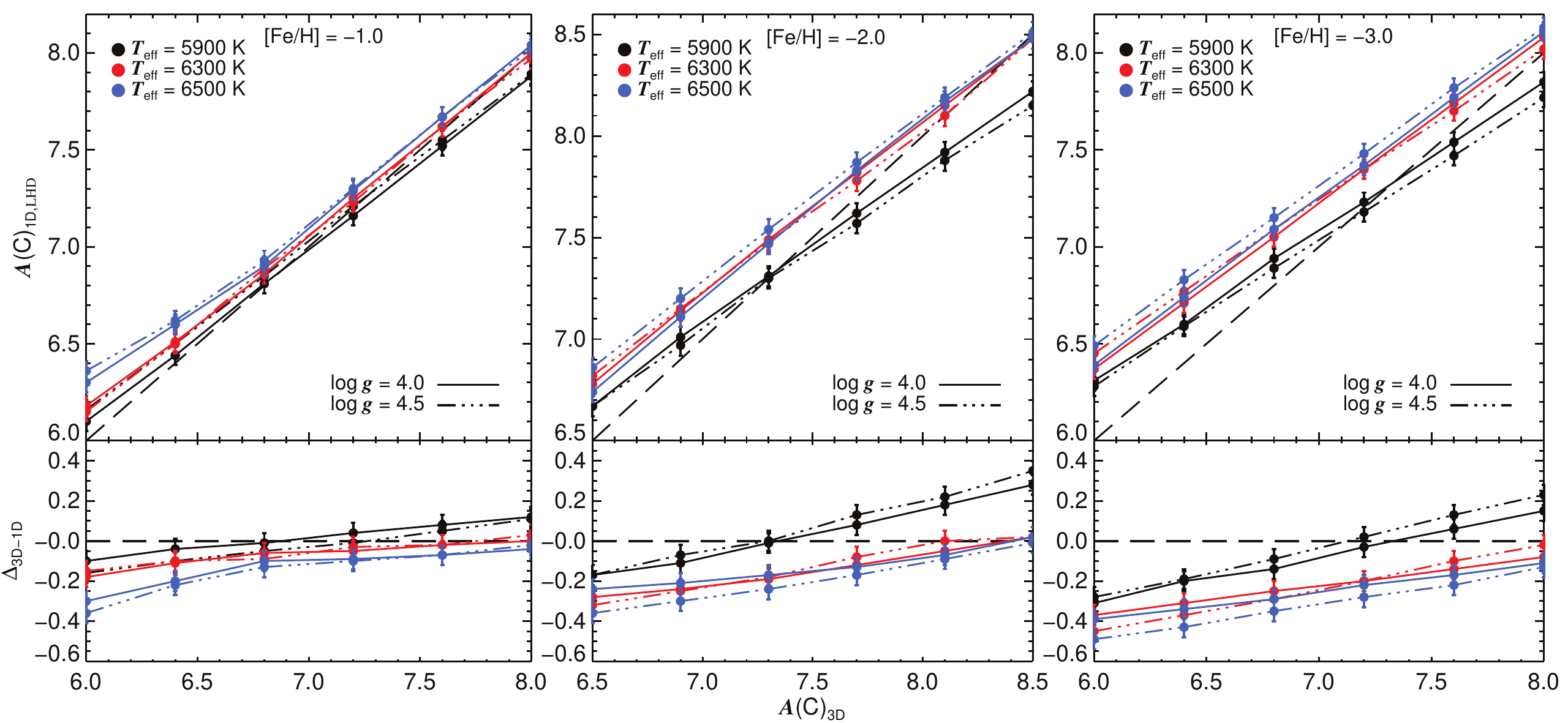}
	\caption{The $\feh=-1.0$ (left panel), $\feh=-2.0$ (middle panel) $\feh=-3.0$ (right panel) abundance corrections from Table~\ref{tab:3Dcorrections}, where we have fixed the C/N and C/O ratios. The top panels presents absolute $\acod$ as a function of $A({\rm C})_{\rm 3D}$ and the bottom panels provide $\Delta_{\rm 3D - 1D}$ as a function of $A({\rm C})_{\rm 3D}$. The dashed line represents the condition $A({\rm C})_{\rm 3D}=\acod$.}
	\label{fig:acorrections}
\end{center}
\end{figure*}

Table~\ref{tab:3Dcorrections} provides the results from fitting the 1D \odx\ synthesis to the 3D synthesis -- using the $\chi^2$ test described above -- in the form of an abundance correction, $\Delta_{\rm 3D - 1D} = A({\rm C})_{\rm 3D} - \acod$. (The $\feh=-2.0$ CNO abundance grid is slightly larger for reasons given in Sect.~\ref{sec:abundances}.) Every abundance correction given assumes that the 3D oxygen abundance is scaled with the carbon abundance, such that ${\rm C/O} = 0.21$. Fitting the entire G-band is in some cases very difficult to do, and as such we had to decide where to fit the band during our $\chi^2$ test. When the band head in the 3D profile saturated, the tail of the G-band was used, and when the tail was too weak, the band head or the entire G-band was used. This presents an unavoidable bias on our corrections, altering $\Delta_{\rm 3D - 1D}$ by $\sim0.02-0.03$\,dex. We also find that our $\chi^2$ tests show further deviations in abundance due to the change in the 1D \odx\ broadening. $\acod$ could shift by $<0.02$\,dex when the macroturbulence is fixed compared to when it is allowed to vary as a free parameter. This is because we treat the broadening of the 1D \odx\ profile in a global way, assuming a single velocity component value across the G-band. We have seen in Sect.~\ref{sec:gbandformation} that individual lines form throughout the atmosphere within different velocity fields. Due to its dynamic nature, the velocity fields are properly treated in the 3D atmosphere. This means that the line profiles of the 1D \odx\ and 3D synthesis will differ, in some cases by a notable amount. Therefore, we expect that the $\chi^2$ test will find a value for the 1D \odx\ broadening that, on average, satisfies the minimum, but because the method of global broadening is imperfect the abundance found will have an uncertainty associated with it. As such we cannot claim accuracies on $\Delta_{\rm 3D - 1D}$ better than $0.04$\,dex. Erring on the side of caution, we do not claim abundance correction accuracies of less than $0.05$\,dex.

Fig.~\ref{fig:acorrections} presents $\acod$ (top panels) and $\Delta_{\rm 3D - 1D}$ (bottom panels) as a function of $A({\rm C})_{\rm 3D}$ from the data given in Table~\ref{tab:3Dcorrections}. It is shown that for any metallicity, temperature and gravity, $\Delta_{\rm 3D - 1D}$ is negative at low $A({\rm C})_{\rm 3D}$, making $\acod$ larger than $A({\rm C})_{\rm 3D}$, but as $A({\rm C})_{\rm 3D}$ increases, $A({\rm C})_{\rm 3D}\rightarrow\acod$, except when $\teff\approx5900$\,K as we see $\acod<A({\rm C})_{\rm 3D}$, making $\Delta_{\rm 3D - 1D}$ positive.

\subsection{Impact of the 3D computations on the carbon abundance in extremely metal-poor and CEMP stars} 

At metallicities around $\feh=-3.0$, different types of extremely metal-poor (EMP) stars are found: classical metal-poor stars with ${\rm [C/Fe]} < +1.0$ ($\sim70\%$ of the stars at this metallicity) and carbon-enhanced metal-poor stars with ${\rm [C/Fe]} > +1.0$. 

\subsubsection{Classical metal-poor stars}

It has been found (under 1D) that the classical turn-off metal-poor stars have a ratio ${\rm [C/Fe]}=+0.45 \pm 0.10$ \citep{Bonifacio2009} with ${\rm [O/Fe]}\approx0.7$\,dex, i.e. $A({\rm C})=5.95$  and $A({\rm O})=6.5$. These values are close to the values given for the C and O abundances in the first line of Table~\ref{tab:3Dcorrections} at $\feh=-3.0$. We deduce that for a mean temperature of 6300\,K and a gravity $\logg=4.0$ the 3D-1D correction, $\Delta_{\rm 3D - 1D}$, is close to $-0.4$\,dex. As a consequence the abundance of carbon in the classical EMP stars would be ${\rm [C/Fe]} \simeq +0.05$\,dex. 

\subsubsection{Carbon-rich metal-poor stars}

At a metallicity of about $\feh=-3.0$ two kinds of carbon-rich stars are observed \citep[e.g.][]{Spite2013,Bonifacio2015}, stars with a carbon abundance $A({\rm C})$ of about 6.8 (under 1D) that are most likely form from a gas cloud with this carbon abundance, and stars with a very high carbon abundance, close to the solar value ($A({\rm C}) \simeq 8.3$) where the atmosphere has been enriched by the ejecta of an AGB companion (now a white dwarf). From Table~\ref{tab:3Dcorrections}, for the stars on the low band of the carbon abundance we expect $\Delta_{\rm 3D - 1D}\approx-0.3$\,dex, but for the stars on the higher plateau at a temperature of $6300$\,K the correction is practically null.

As a consequence the mean value of the carbon abundance on the lower ``plateau'' should be close to $6.2$ and the difference between the classical metal-poor stars and the carbon-rich metal-poor stars would increase by about $0.1$\,dex. The abundance of carbon of the stars on the higher carbon band would not change. 

\subsection{Impact of oxygen on the 3D abundance correction}
\label{sec:oxygen}

\begin{figure*}[ht]
	\includegraphics[scale=1]{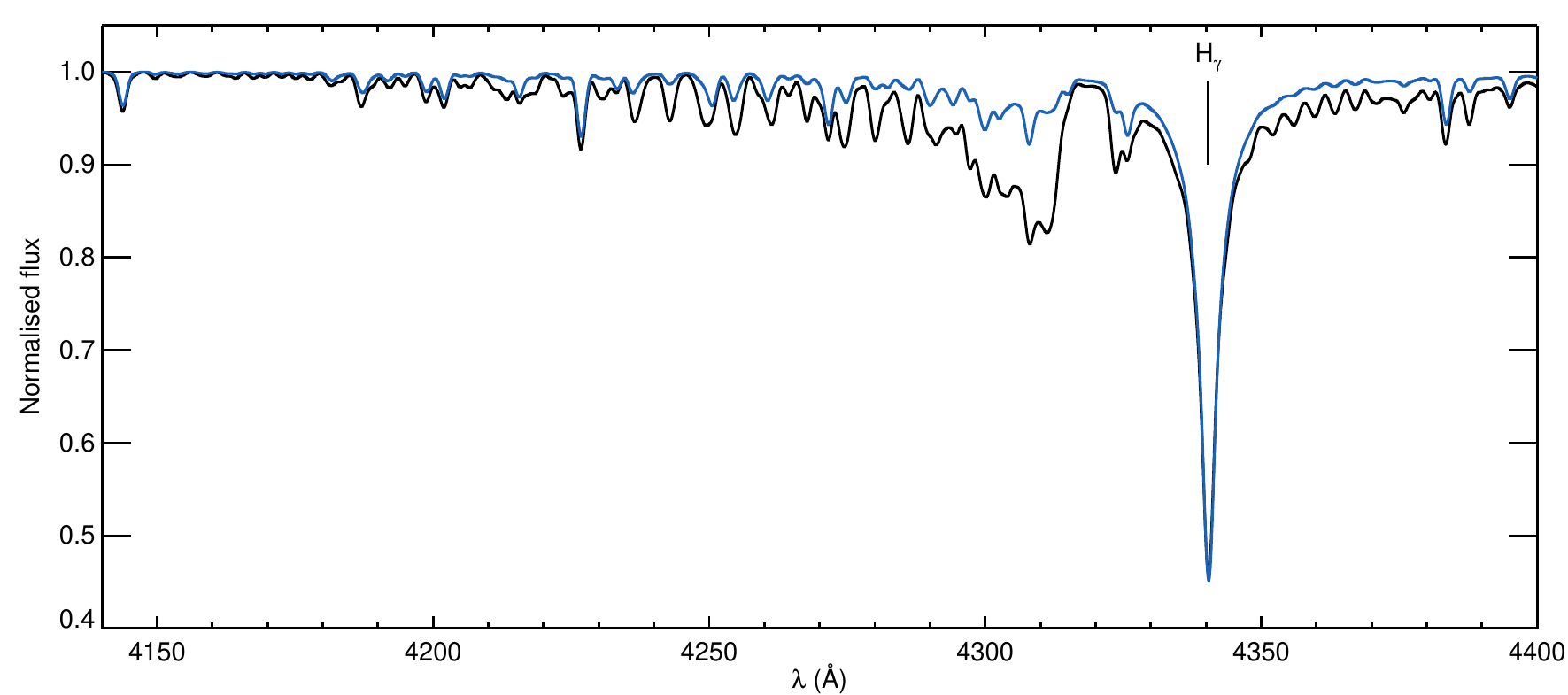}
	\caption{The G-band for a 3D synthesis (black line) and a 1D \odx\ synthesis (blue line), which demonstrates the effect $A({\rm O})$ has on the 3D synthesis. Both synthetic bands have the same CN abundances as Fig.~\ref{fig:gband_1d3d}, $A({\rm C})=6.80$ and $A({\rm N})=6.21$, but $A({\rm O})$ has been decreased relative to $A({\rm C})$ so that $A({\rm O})=6.20$, implying that ${\rm C/O}=3.98$.}
	\label{fig:gband_oeffect}
\end{figure*}

To investigate the effect the oxygen abundance, and hence the C/O ratio, has on the 3D carbon abundance correction, $\Delta_{\rm 3D - 1D}$, we computed an additional synthetic G-band with an increased C/O ratio (3.98) using every atmosphere listed in Table~\ref{tab:atmgrid}. Fig.~\ref{fig:gband_oeffect} demonstrates the effect that enhancing the C/O ratio from 0.21 (Fig.~\ref{fig:gband_1d3d}) to $3.98$ has on the G-band when $A({\rm C})$ is fixed. It is immediately clear that the depletion of oxygen leads to an enhancement in the strength of the CH transitions in 3D, while the strength of the 1D synthesis remains constant.

Stars with a high C/O ratio are known to exist \citep[][their Table~4]{Sivarani2006}, so testing an abundance configuration with ${\rm C/O}=3.98$ is a useful exercise, considering the influence oxygen has on the G-band (Sect.~\ref{sec:gbandformation}). Table~\ref{tab:cocorrections} lists the additional correction found when $A({\rm C})_{\rm 3D}$ is fixed and C/O is increased. For the $\feh=-2.0$ and $\feh=-3.0$ model atmospheres, it is very clear that not only is the C/O ratio particularly important, but it has a larger influence as the temperature and gravity increase.

\begin{table*}[ht]
\begin{center}
\caption{Abundance corrections found from fixing $A(\rm C)$ and varying the C/O ratio between $0.21$ and $3.98$. The errors associated with the reported values remain $\pm0.05$\,dex.}
\begin{tabular}{l c c c c c c c c c c r}
\hline\hline
$A({\rm C})_{\rm 3D}$ & $A({\rm N})_{\rm 3D}$ & $A({\rm O})_{\rm 3D}$ & C/O & \multicolumn{8}{c}{$\ctd$} \\
\hline
&&&&\multicolumn{2}{c}{$\teff\approx5900$} && \multicolumn{2}{c}{$\teff\approx6300$} && \multicolumn{2}{c}{$\teff\approx6500$} \\
&&&& $\logg=4.0$ & $\logg=4.5$ && $\logg=4.0$ & $\logg=4.5$ && $\logg=4.0$ & $\logg=4.5$ \\
\cline{5-6} \cline{8-9} \cline{11-12} \vspace{-0.8em} \\
\hline
\multicolumn{12}{c}{$\feh=-1.0$}\\
\hline
$6.80$ & $6.21$ & $6.20$ & $3.98$ & $-0.06$ & $-0.15$ && $-0.08$ & $-0.15$ && $-0.14$ & $-0.20$ \\
$6.80$ & $6.21$ & $7.47$ & $0.21$ & $-0.01$ & $-0.05$ && $-0.06$ & $-0.09$ && $-0.10$ & $-0.13$ \\
\hline
\multicolumn{12}{c}{$\feh=-2.0$}\\
\hline
$7.30$ & $6.71$ & $6.70$ & $3.98$ & $-0.41$ & $-0.53$ && $-0.51$ & $-0.66$ && $-0.62$ & $-0.72$ \\
$7.30$ & $6.71$ & $7.97$ & $0.21$ & $-0.01$ & $+0.00$ && $-0.19$ & $-0.19$ && $-0.17$ & $-0.24$ \\
\hline
\multicolumn{12}{c}{$\feh=-3.0$}\\
\hline
$6.80$ & $6.21$ & $6.20$ & $3.98$ & $-0.66$ & $-0.67$ && $-0.64$ & $-0.85$ && $-0.88$ & $-0.92$ \\
$6.80$ & $6.21$ & $7.47$ & $0.21$ & $-0.14$ & $-0.09$ && $-0.25$ & $-0.29$ && $-0.29$ & $-0.35$ \\
\hline
\label{tab:cocorrections}
\end{tabular}
\end{center}
\end{table*}

While the data in Table~\ref{tab:cocorrections} provides very useful indicators of how much the oxygen abundance influences the G-band, the fact that a single value is reported allows us to make very few postulates on how changing the C/O ratio affects the G-band at various values of $A({\rm C})_{\rm 3D}$. As such, we computed additional synthesis for a single atmosphere, d3t63g40mm30n02, where we fixed the oxygen abundance while increasing the carbon abundance, thus increasing the C/O ratio. The data is tabulated in Table~\ref{tab:d3t63g40mm30n02_co}, along with the companion data for this atmosphere from Table~\ref{tab:3Dcorrections}.

\begin{table}[!th]
\begin{center}
\caption{An expanded test of the impact that the C/O ratio has on the abundance correction for the atmosphere d3t63g40mm30n02.}
\begin{tabular}{r c c r}
\hline\hline
C/O & $\actd$ & $A({\rm O})_{\rm 3D}$ & $\ctd$ \\
\hline
\noalign{\vskip 0.4em}
$0.21$ & $6.00$ & $6.67$ & $-0.37$ \\
$0.21$ & $6.40$ & $7.07$ & $-0.31$ \\
$0.21$ & $6.80$ & $7.47$ & $-0.25$ \\
$0.21$ & $7.20$ & $7.87$ & $-0.20$ \\
$0.21$ & $7.60$ & $8.27$ & $-0.14$ \\
$0.21$ & $8.00$ & $8.67$ & $-0.08$ \\
\noalign{\vskip 0.4em}
$0.63$ & $6.00$ & $6.20$ & $-0.44$ \\
$1.58$ & $6.40$ & $6.20$ & $-0.57$ \\
$3.98$ & $6.80$ & $6.20$ & $-0.64$ \\
$10.00$ & $7.20$ & $6.20$ & $-0.63$ \\
$25.12$ & $7.60$ & $6.20$ & $-0.58$ \\
$63.10$ & $8.00$ & $6.20$ & $-0.52$ \\
\hline
\label{tab:d3t63g40mm30n02_co}
\end{tabular}
\end{center}
\end{table}

\begin{figure}[!th]
	\begin{center}
	\includegraphics[scale=0.5]{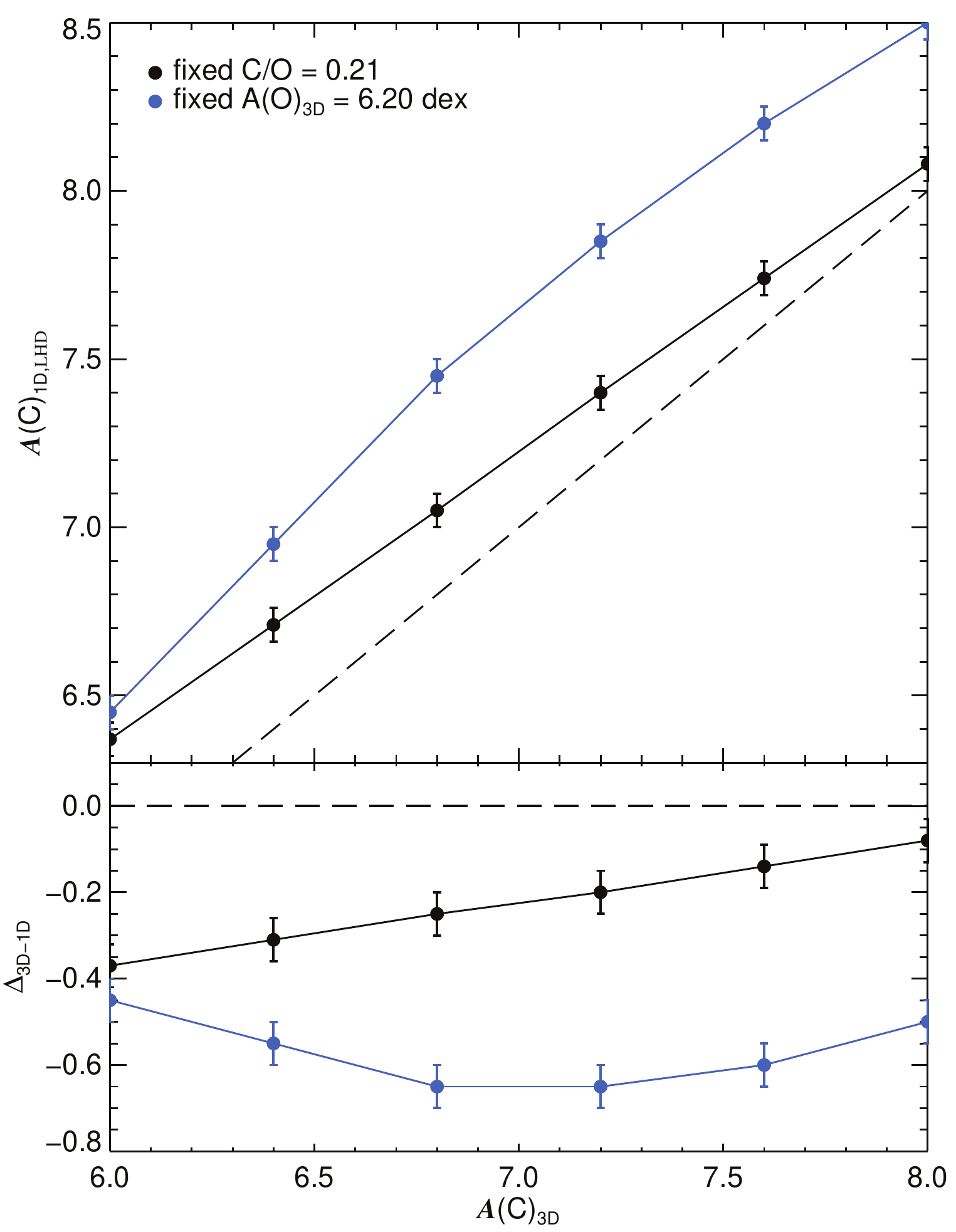}
	\caption{A graphical representation of values tabulated in Table~\ref{tab:d3t63g40mm30n02_co}, which demonstrate the effect of increasing the C/O ratio on $A({\rm C})_{\rm 3D}$ (blue symbols) compared with when C/O is fixed at $0.21$ (black symbols). The dashed line represents the condition $A({\rm C})_{\rm 3D}=\acod$.}
	\label{fig:coratio}
	\end{center}
\end{figure}

It would appear that the increase in the C/O ratio with increasing $A({\rm C})_{\rm 3D}$ further enhances the correction needed. However, as $A({\rm C})_{\rm 3D}>6.80$\,dex (and ${\rm C/O}>3.98$), $\delta\Delta_{\rm 3D - 1D}/\delta A({\rm C})_{\rm 3D}$ starts to decrease. Fig.~\ref{fig:coratio} demonstrates this effect well. After checking the contribution functions of the band head ($4280-4310$\,\AA) it was found that the increasing C/O forces the formation band head to higher layers in the atmosphere until an increasing amount the band head begins to form outside the computational box when $3.98<{\rm C/O}\leq10.00$ and $6.80<A({\rm C})_{\rm 3D}\leq7.20$, making the abundances, and hence $\Delta_{\rm 3D - 1D}$, less reliable. We stress that the band head was not used to determine $\Delta_{\rm 3D - 1D}$ at high values of $A({\rm C})_{\rm 3D}$ because the saturation of the feature made fitting the 1D \odx\ spectra very difficult. However, the fact that we start to see formation outside the computational box no doubt makes the $\Delta_{\rm 3D - 1D}$ reported in Table~\ref{tab:d3t63g40mm30n02_co} for the higher C/O ratios and $A({\rm C})_{\rm 3D}$, and the change in gradient seen in Fig.~\ref{fig:coratio}, questionable results. 

This suggests that there is a limit on how far the CNO abundances can be varied in the 3D synthesis we have computed using the models from the CIFIST grid and provides an initial warning on certain limitations of 3D LTE synthesis used to analyse the G-band from these model atmospheres. A similar issue was previously noted in \citet{Bonifacio2013} when fitting the CN-band. Two solutions could address this problem:

\begin{enumerate}
\item The outer regions of the CIFIST model grid need be extended so that those lines that currently form outside the computational box are fully considered. However, chromospheric effects would need to be considered.

\item An NLTE treatment of the CH transitions. This would mean that certain stars with high C/O ratios, such as some of those presented in \citet{Sivarani2006}, cannot be analysed using a 3D G-band grid under the assumption of LTE.
\end{enumerate}

\section{Discussion}
\label{sec:discussion}

We have seen that in 3D the oxygen abundance has a strong effect on the CH transitions that make up the G-band. Ideally, one would know the oxygen abundance of a star so that it can be fixed in the spectral synthesis of the G-band, and the true carbon abundance can be attained from measurements of the G-band. Unfortunately, it is not always possible to measure atomic oxygen lines in a metal-poor star and one must look at molecular features (if available), like the OH molecular features in the UV. From Figs.~\ref{fig:levpop} \&~\ref{fig:contfunc} it is quite clear (and also expected) that there is a direct relationship between the OH and CO molecules as the oxygen abundance is altered, but what is unclear is whether they have a similar relationship as the one shared between the CH and CO molecules (demonstrated in Sects.~\ref{sec:gbandformation}~\&~\ref{sec:gband_analysis}), when the carbon abundance is changed. \citet{Dobrovolskas2013} state that non-carbon bearing molecules, like OH, are not affected by CO like CH is, but they do not report on the effects of varying the carbon or oxygen abundances have on the OH and CH molecules, respectively. Rather, they examine the effects that CO have for varying temperature and saw no evidence that OH is affected by CO like CH is. \citet{Collet2007}, however, demonstrated that the C/O ratio is important to CH and OH, due to their interrelationship with CO. \citet{Wedemeyer2005} published an extensive review on the carbon and oxygen reaction network implemented in \cobold\ (see their Table~1) and examined the role of CO in the solar atmosphere in great detail. They show that a complex interconnecting relationship exists between the CH, OH and CO molecules (as seen in Fig.~\ref{fig:levpop}). Therefore, changing the carbon and oxygen abundances alters the ratio of available carbon and oxygen atoms, impacting the relationship between CH, CO and OH formation. This means that there is a relationship between the carbon abundance and the formation of OH, which would mean that the C/O ratio not only affects CH abundances, but also OH abundances in a 3D atmosphere.

To examine this further, we conducted a small experiment using a very small number of individual OH and CH lines found in the UV. We synthesised the OH lines using a fixed oxygen abundance and the CH lines using a fixed carbon abundance. We then varied the C/O ratio so that $0.1\leq{\rm C/O}\leq10$. Further details of this and an example UV CH and OH line are presented in Appendix~\ref{appdx:COratio}. We examined this using the same turn-off dwarf star atmospheres tested in Sect.~\ref{sec:gbandformation}, i.e. d3t63g40mm30n02. We measured the resultant equivalent widths of the synthetic spectra and found that the OH molecules do show a dependency on the carbon abundance inferred by the C/O ratio in a metal-poor 3D model atmosphere, and that their dependency reduces as the metallicity increases, as we see in CH. This is expected given that we see the relationship between the CH, CO and OH molecules become less sensitive to changing $A({\rm C})$ (and $A({\rm O})$  as shown in Figs.~\ref{fig:levpop}, \ref{afig:levpop1} and \ref{afig:levpop2}). This means that the C/O ratio is an equally important parameter to CH and OH in 3D.

When we compare the CH and OH synthesis in 3D with their counterpart 1D synthesis (like in Fig.~\ref{afig:choh}), we find that C/O has a peculiar effect on the 3D corrections. We find that the 3D CH corrections tend towards zero as the C/O ratio decreases, and the 3D OH corrections tend towards zero as the C/O ratio increases. This demonstrates a clear anti-correlated interrelationship between oxygen and carbon in a 3D atmosphere through their molecular species, meaning that the carbon and oxygen abundances must be simultaneously measured in a 3D spectrum to attain the true 3D correction. A more complete examination of this effect is currently under way.

\section{Conclusions}
\label{sec:conclusions}

We present the first results of a full 3D treatment of the G-band under the assumption of LTE. As such we have discovered several unexpected results, previously unseen in the classical 1D LTE (or NLTE) paradigm.

\begin{enumerate}

\item The abundance of oxygen has a strong influence on the G-band and its impact to the abundance correction, $\Delta_{\rm 3D - 1D}$, varies with $A({\rm C})_{\rm 3D}$ such that the C/O ratio becomes an important parameter under 3D.

\item As it is difficult to measure the oxygen abundance in a metal-poor atmosphere, most metal-poor stars have an unknown oxygen abundance. This implies that the C/O ratio becomes an additional variable in 3D. Therefore, to properly fit the G-band the carbon abundance and C/O ratios must be explored simultaneously. This yields both a carbon abundance and a limit on the oxygen abundance

\item The impact of the C/O ratio is due to the importance of the CO molecule to CH and OH molecules in a 3D atmosphere. The durability of CO and the low dissociation energies of CH and OH combined with the cooler temperatures and large temperature fluctuations inherent in metal-poor 3D atmospheres dramatically increases the formation of CO at the expense of CH and OH. This effect weakens in the more metal-rich model atmospheres, as the temperatures become hotter, and the C/O ratio (and hence the oxygen abundance) becomes less important. 

\item The C/O ratio is not an important parameter to CH or OH formation under 1D or \atd\ at any metallicity tested. As the 1D \odx\ and \atd\ model atmospheres do not reach the low temperatures found in the low-temperature fluctuations of the 3D atmosphere, CO molecules do not dominate over OH and CH, and the C/O ratio does not impact the CH and OH molecule formation, and hence their line strengths.

\item As both the C/O ratio and $A({\rm C})_{\rm 3D}$ are increased, parts of the G-band -- such as the band head -- form outside the confines of the 3D model atmosphere, meaning that reported abundances start to become unreliable. A solution to this is to either explore 3D CH formation under the assumption of NLTE, however, this is not currently possible, or extend the outer regions of the CIFIST model atmospheres. One must then consider chromospheric effects.

\item Not only is the C/O ratio an important parameter in the G-band (and other CH transitions), it is also an important parameter to OH transitions.

\item Not surprisingly, the C/O ratio has the opposite effect on the strengths of the CH and OH transitions; when C/O gets smaller, the OH lines become stronger while the CH transitions become weaker. This demonstrates an interconnecting behaviour between the CH and OH molecules through their affiliation with CO.

\item The 3D corrections presented in Table~\ref{tab:3Dcorrections} and Fig.~\ref{fig:acorrections} imply that the carbon abundances would decrease in stars on the low carbon plateau, but would be mostly unchanged in stars on the high carbon plateau.

\end{enumerate}

The results found in this work serve as a guide for future works on the G-band constructed using 3D model atmosphere and spectrum synthesis. The logical next step to this work is to use the grid we have published here to analyse real stellar data, which we are currently undertaking.

\begin{acknowledgements}

This project is funded by FONDATION MERAC and the matching fund granted by the Scientific Council of Observatoire de Paris. We acknowledge support from the Programme National de Cosmologie et Galaxies (PNCG) and Programme  National de Physique Stellaire (PNPS) of the Institut National de Sciences de l'Univers of CNRS.
This work was granted access to the HPC resources of MesoPSL financed by the Region Ile de France and the project Equip@Meso (reference ANR-10-EQPX-29-01) of the programme Investissements d’Avenir supervised by the Agence Nationale pour la Recherche.
This work was supported by Sonderforschungsbereich SFB 881 "The Milky Way System" (subproject A4) of the German Research Foundation (DFG).

\end{acknowledgements}

\nocite{*}
\bibliographystyle{aa}
\bibliography{28602}

\onecolumn

\begin{appendix}

\section{OH and CH sensitivities to the C/O ratio}
\label{appdx:COratio}

\begin{figure*}[!th]
	\begin{center}
	\includegraphics[scale=0.458]{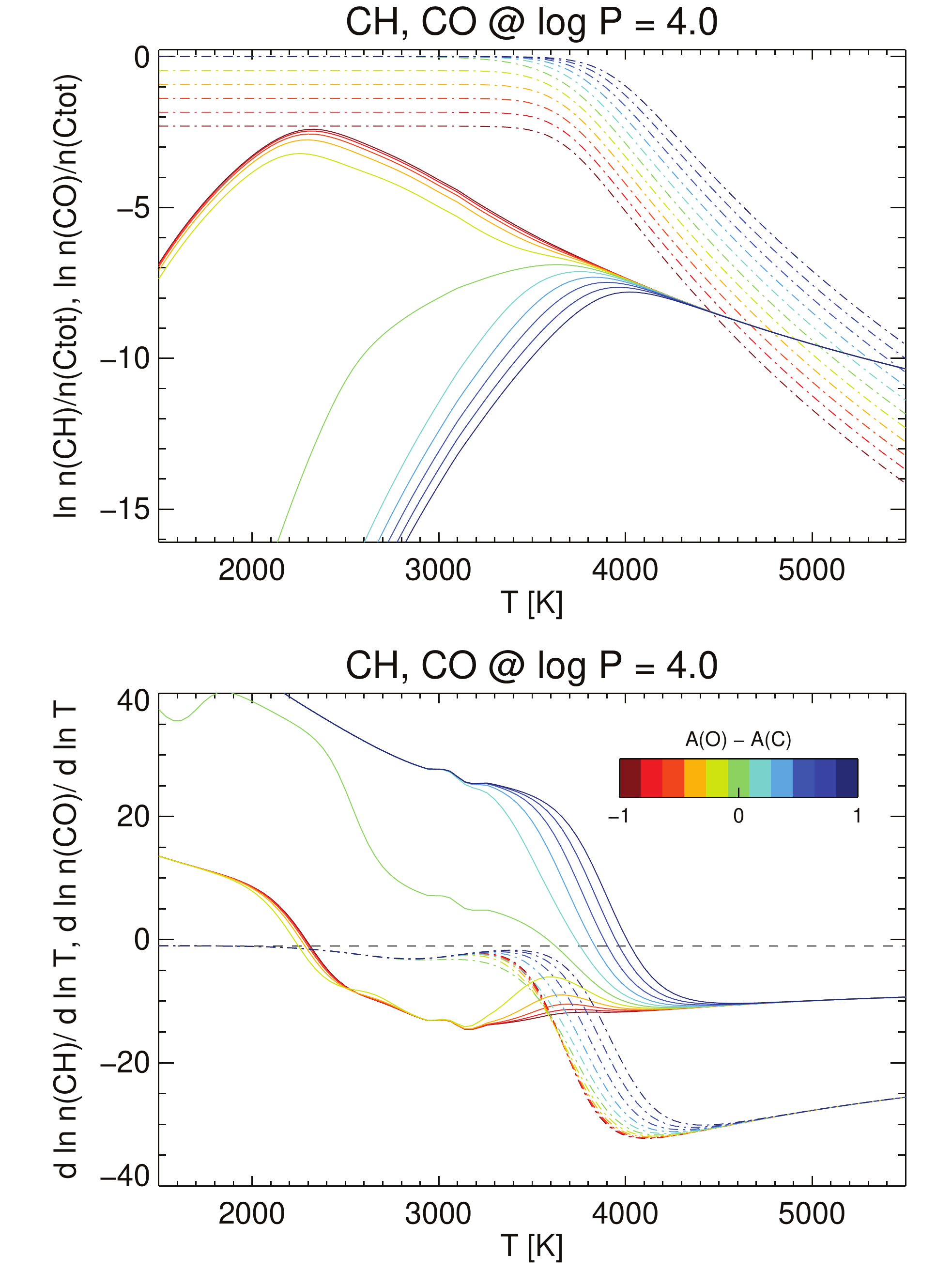}
	\includegraphics[scale=0.458]{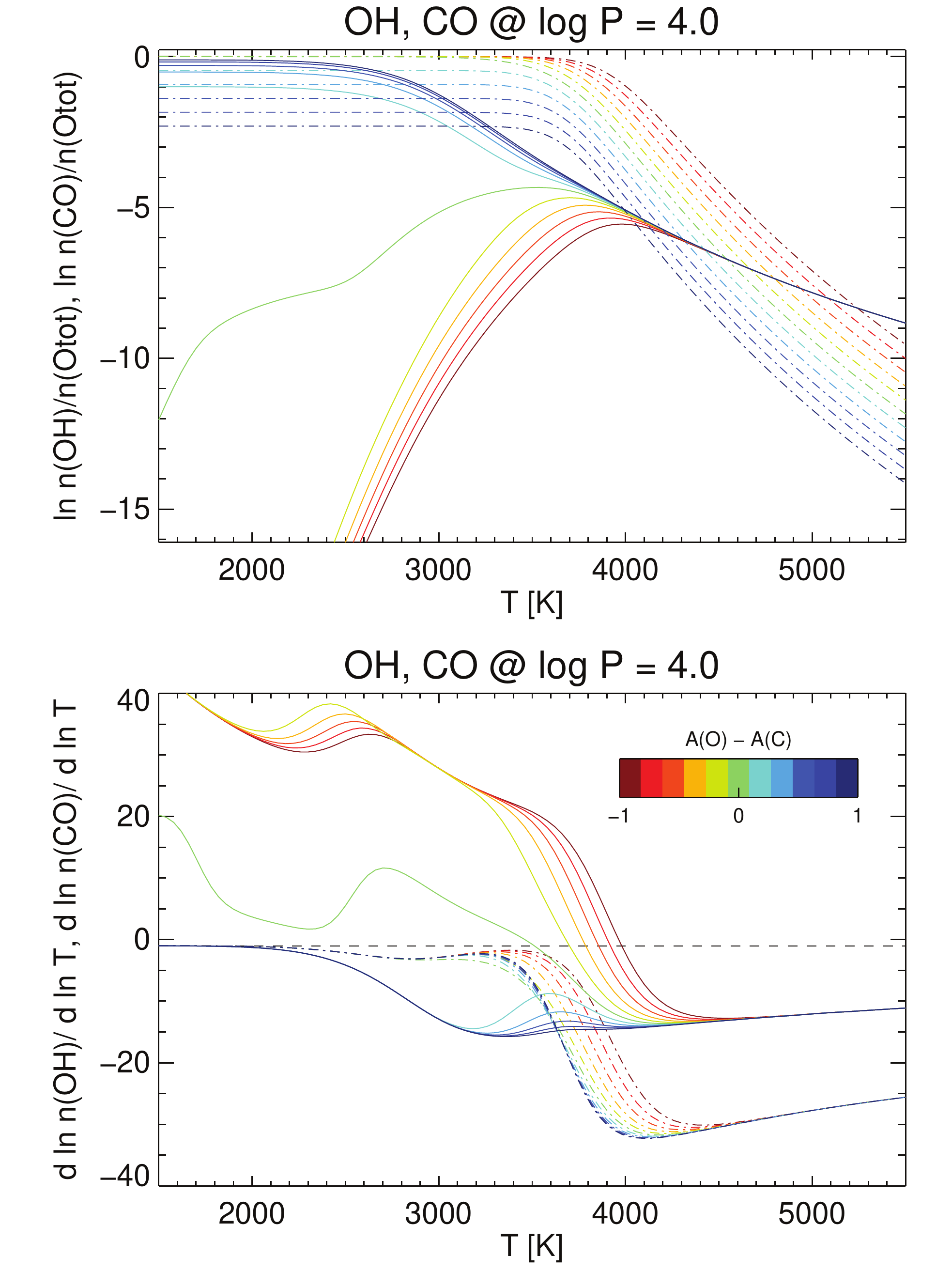}
	\caption{\textit{Top panels}: The number densities of carbon atoms in CH (solid lines) and CO (dashed-dot lines), relative to the total number of carbon atoms (left panel), and the number densities of oxygen atoms in OH (solid lines) and CO (dashed-dot lines), relative to the total number of oxygen atoms (right panel). 
	\textit{Bottom panels}: CH (solid lines) and CO (dashed-dot lines) temperature sensitivities (left panel), and OH (solid lines) and CO (dashed-dot lines) temperature sensitivities (right panel). Figures depicting CH (left panels) have been computed assuming $A({\rm C})=6.0$ with $5.0\leq A({\rm O})\leq7.0$. Figures depicting OH (right panels) have been computed assuming $A({\rm O})=6.0$ with $5.0\leq A({\rm C})\leq 7.0$.}
	\label{afig:cnumberdensities}
	\end{center}
\end{figure*}

We present some further details of the test described in Sect.~\ref{sec:discussion}. Fig.~\ref{afig:cnumberdensities} demonstrates the effects the C/O ratio have on CH, CO and OH at different temperatures. For simplicity, we have fixed the gas pressure at $10^4\,[{\rm cgs}]$, which is typical for a metal-poor dwarf star. Number densities of carbon and oxygen atoms in CH, OH and CO molecules, and CH, OH and CO temperature sensitivities were then computed for varying C/O ratios (i.e. $0.1\leq{\rm C/O}\leq10$, blue to red respectively). While we do not expect these behaviours to be fully reproduced in a full 3D model atmosphere, it does serve to clearly illustrate the C/O ratio dependence on the CH and OH molecules and their interrelationship.

The figure shows that the effect the C/O ratio has on the CH and OH molecules is highly temperature dependent. It is clear from the top two panels of this figure that as the C/O ratio is varied, several things happen:

\begin{enumerate}

\item The carbon atoms in the CO molecules show a lower sensitivity to C/O at any temperature, relative those in the CH molecules. This demonstrates the durability of this molecule in dwarf star atmospheres due to its high dissociation energy ($11.092$\,eV).

\item It is clear that CH and OH are anti-correlated except in cases were ${\rm C/O}=1.0$.  The higher number of carbon atoms relative to oxygen in regimes where ${\rm C/O}>1.0$ allows for far greater numbers of CH to form at the  expense of OH as temperatures decrease. Greater numbers of OH form at the expense of CH in regimes were ${\rm C/O}<1.0$ as temperatures decrease.

\item ${\rm C/O}=1.0$ (green lines) mark a bifurcation point or boundary between carbon- and oxygen-rich regimes, where the majority of oxygen or carbon atoms are bound to form CO, respectively.

\end{enumerate}

The temperature sensitivities of CH, OH and CO molecules are depicted in the bottom two panels of Fig.~\ref{afig:cnumberdensities}. When the C/O ratio is varied, several things happen: 

\begin{enumerate}

\item The CO molecules show a lower sensitivity to C/O at any temperature, relative to the CH and OH molecules, except for the temperature domain $3.2\leq T \ [10^3\,{\rm K}]\leq 4.4$ where a very slight sensitivity is shown.

\item The temperature domain where all molecules show the highest sensitivity to the C/O ratio is found in the region $3.2\leq T \ [10^3\,{\rm K}]\leq 4.4$.

\item A very high temperature sensitivity is found at ${\rm C/O}=1.0$.

\end{enumerate}

\begin{figure*}[!th]
	\begin{center}
	\includegraphics[scale=0.55]{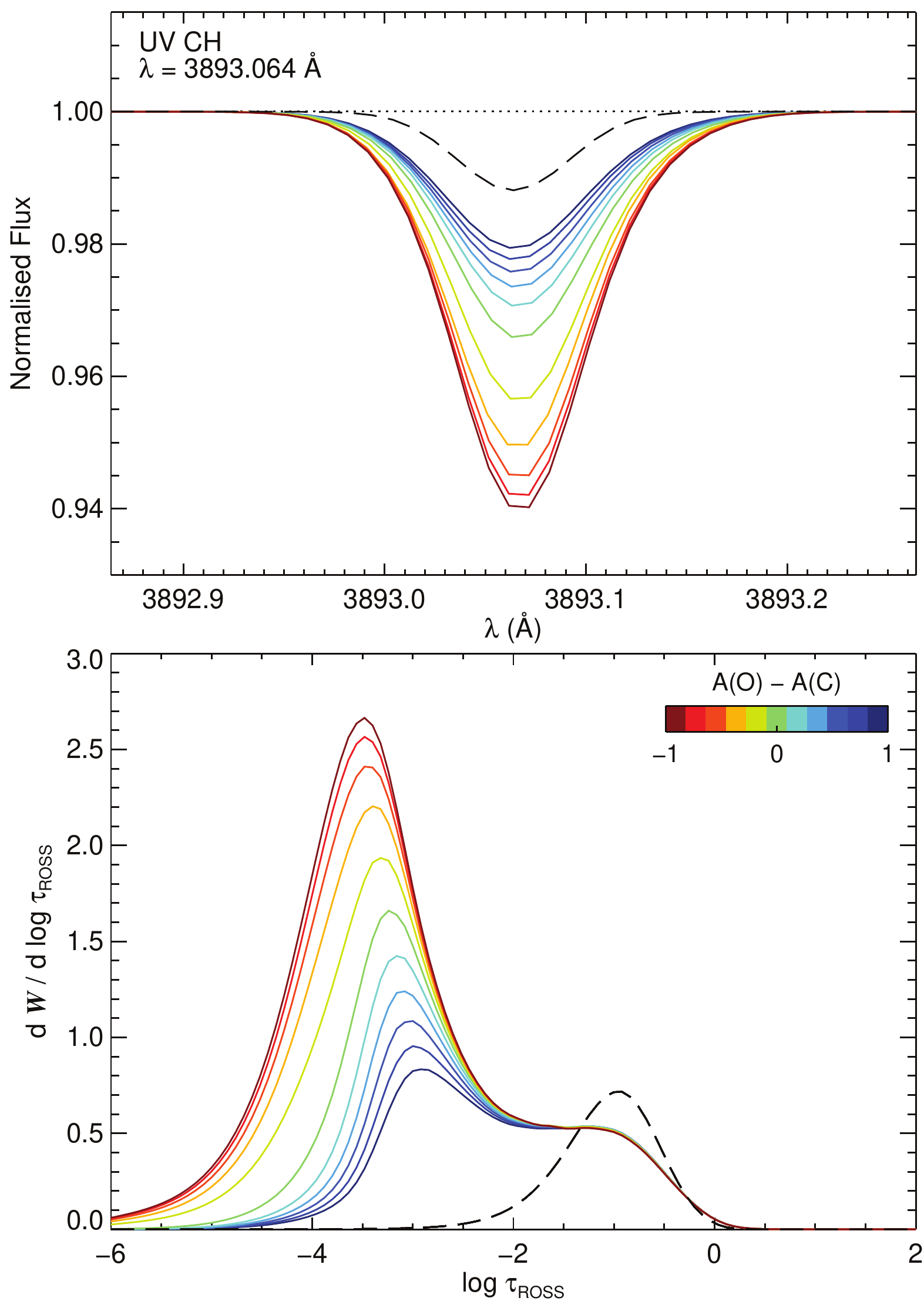}
	\includegraphics[scale=0.55]{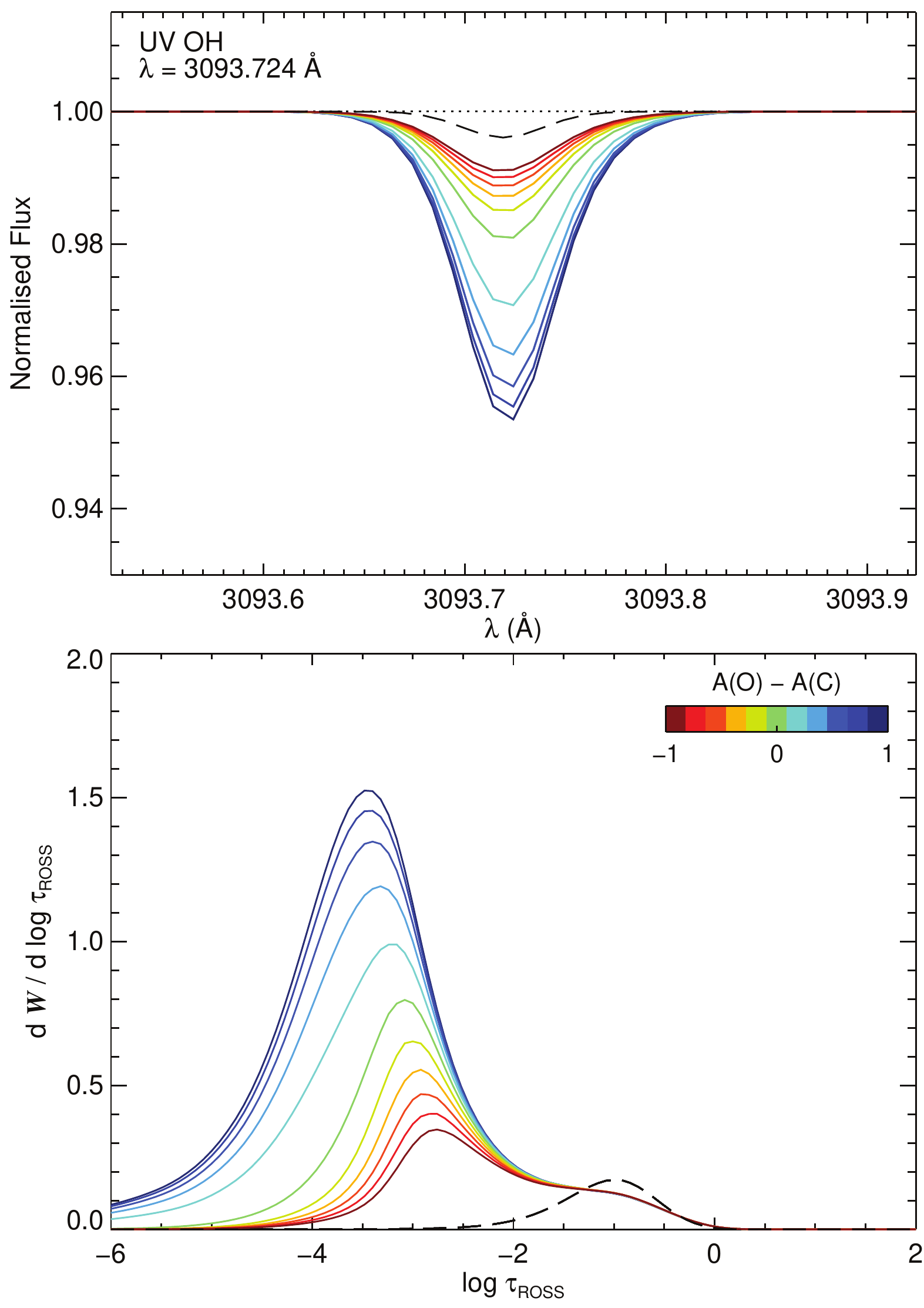}
	\caption{\textit{Left panels}: A CH line (top) computed in full 3D (solid colors) and 1D (black dashed line) and the corresponding equivalent width contribution functions (bottom). \textit{Right panels}: An OH line (top) computed in full 3D (solid colors) and 1D (black dashed line) and the corresponding equivalent width contribution functions (bottom). The abundances and C/O ratios of these lines are identical to those given in Fig.~\ref{afig:cnumberdensities}.}
	\label{afig:choh}
	\end{center}
\end{figure*}

To demonstrate how these sensitivities affect the CH and OH lines, we have provided an example in Fig.~\ref{afig:choh}. For clarity, the color schemes and $A({\rm C})$, $A({\rm O})$, and C/O parameters are identical to those in Fig.~\ref{afig:cnumberdensities}. These lines were computed using the full 3D atmosphere with stellar parameters $\teff/\logg/\feh=6250\,{\rm K}/4.0/-3.0$, i.e. the same used in the figure above, however, we have also included an equivalent synthetic line from 1D \odx\ model atmosphere with identical stellar parameters. We also present the CH and OH line's equivalent width contribution functions. 

The behaviours described above, coupled with the behaviours seen in the contribution functions help to explain some of the properties seen in the spectral lines:

\begin{enumerate}

\item It is clear from the CH and OH line depicted that the C/O ratio has an anti-correlated influence on these lines and that the OH and CH line are equally sensitive to the C/O ratio in this example. 

\item The C/O ratio has very little impact on the 1D CH and OH synthesis. This is because the 1D models do not reach the lower temperatures where the C/O ratio sensitivity is high. This is also the case in the average 3D models. The temperature fluctuations in the full 3D model reach these low temperatures where CO molecular populations dominate and so the C/O ratio becomes important.

\item The 3D contribution functions of the CH lines indicate a high sensitivity to the C/O ratio. As it increases, and the line gets stronger, its formation is pushed towards the outer regions of the 3D atmosphere where temperatures are cooler. When we compare this behaviour with CH number densities in Fig.~\ref{afig:cnumberdensities}, we can see that the CH molecule populations are highest in cooler regions of the atmosphere for cases where ${\rm C/O}>1.0$. As the OH and CH lines show an anti-correlated sensitivity to C/O, the opposite is seen in OH.

\end{enumerate} 

\section{Additional figures}
\label{appdx:afigures}

We present several figures for additional atmospheres that complement the figures presented in Sect.~\ref{sec:gbandformation}. These include identical figures of Figs.~\ref{fig:levpop} and \ref{fig:contfunc} for the $\feh=-1.0$ and $-2.0$ atmospheres. 

\twocolumn
\newpage

\begin{figure*}[!th]
\begin{center}
	\includegraphics[scale=0.713]{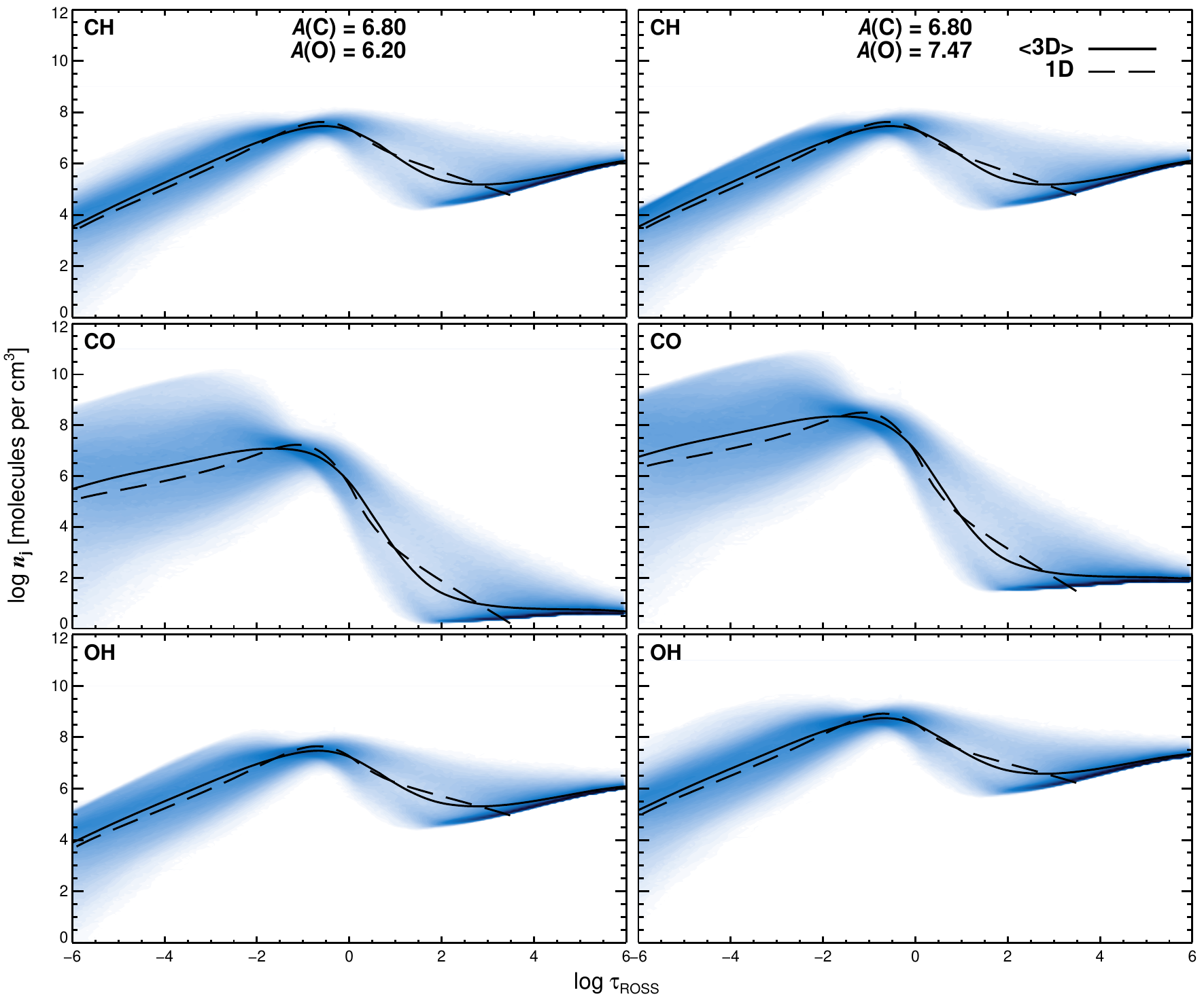}
	\caption{Corresponding $\feh=-1.0$ molecular number density plot as the one presented in Fig.~\ref{fig:levpop}. The large differences on OH, CH and CO formation seen between the 3D and 1D \odx\ have almost completely disappeared.}
	\label{afig:levpop1}
\end{center}
\end{figure*}

\begin{figure*}[!th]
\begin{center}
	\includegraphics[scale=0.713]{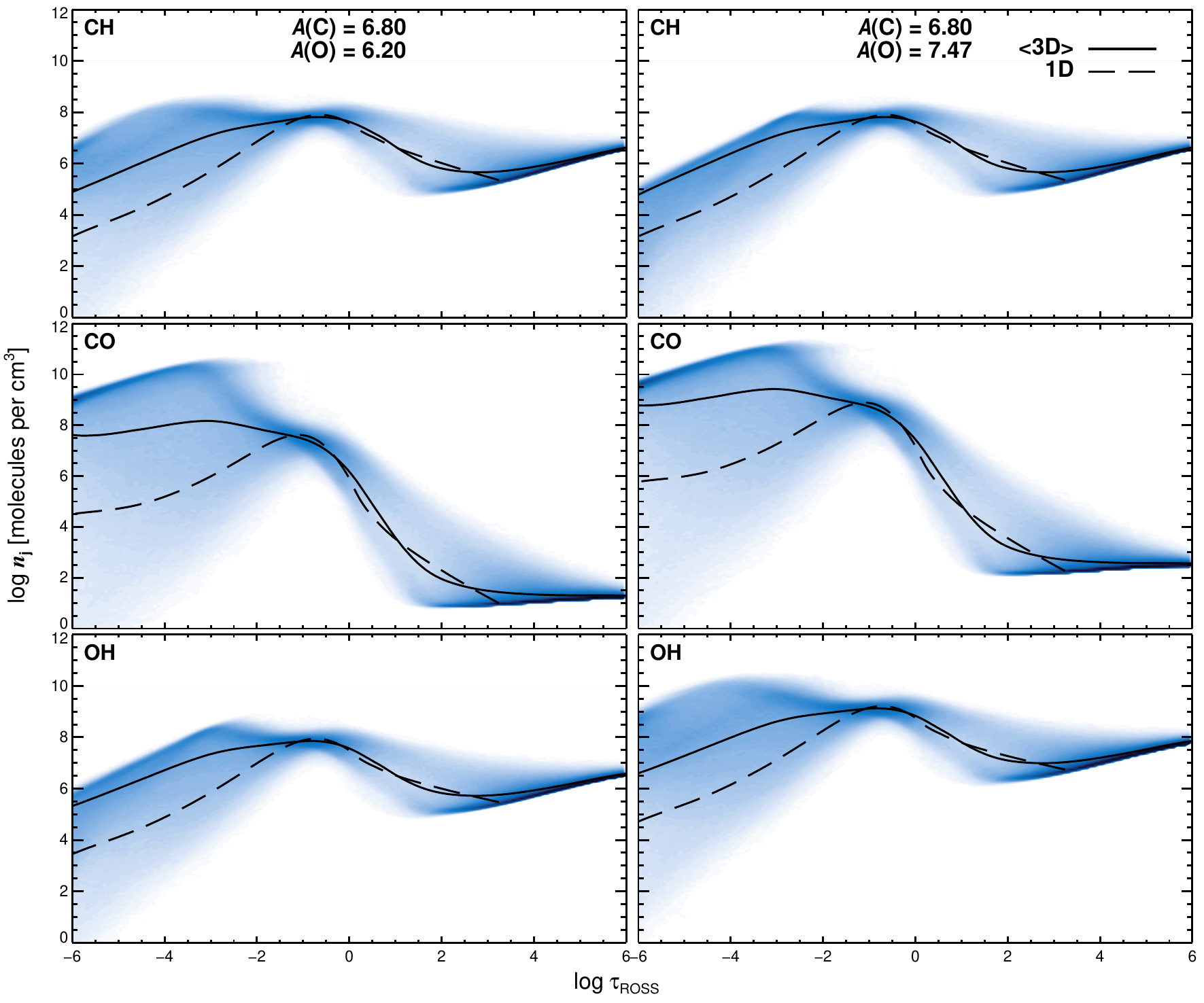}
	\caption{Corresponding $\feh=-2.0$ molecular number density plot as the one presented in Fig.~\ref{fig:levpop}. While the formation of the CH, OH and CO molecules still show significant differences between the 3D and 1D \odx\ model atmospheres, their effect is reduced when compared with those at $\feh=-3.0$ presented in Fig.~\ref{fig:levpop}.}
	\label{afig:levpop2}
\end{center}
\end{figure*}

\begin{figure*}[!th]
	\begin{center}
	\includegraphics[scale=0.877]{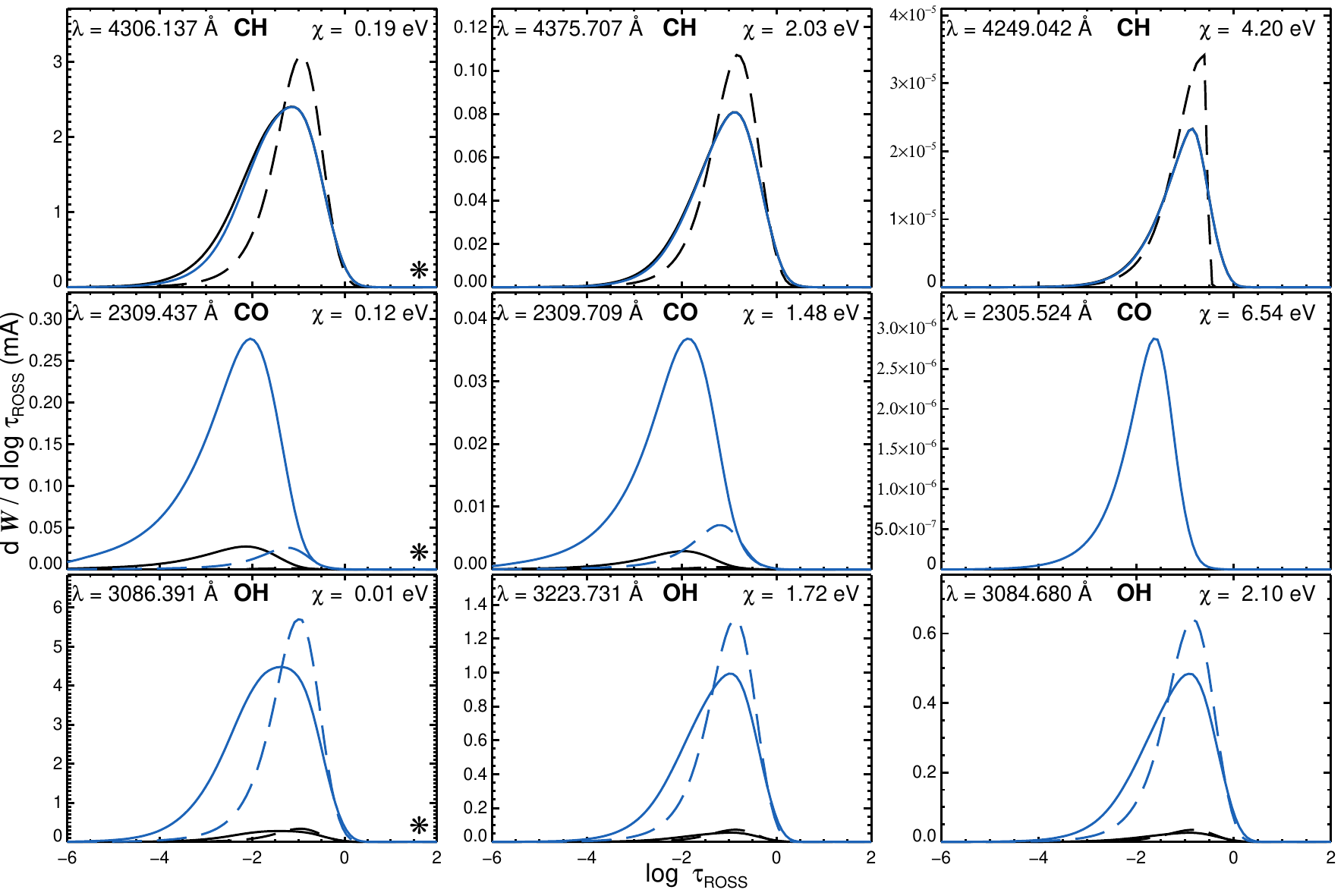}
	\caption{Corresponding $\feh=-1.0$ CH, CO and OH equivalent width contribution functions in 3D (solid lines) and and 1D (dashed lines) calculated for lines with different excitation potentials in the d3t63g40mm10n02 atmosphere. $\actd=\acod= 6.80\,{\rm dex}$ with $\aotd=\aood=6.20\,{\rm dex}$ (black) and $7.47\,{\rm dex}$ (blue). The oxygen abundance does not affect 1D CH contribution function.}
	\label{afig:contfunc1}
	\end{center}
\end{figure*}

\begin{figure*}[!th]
	\begin{center}
	\includegraphics[scale=0.877]{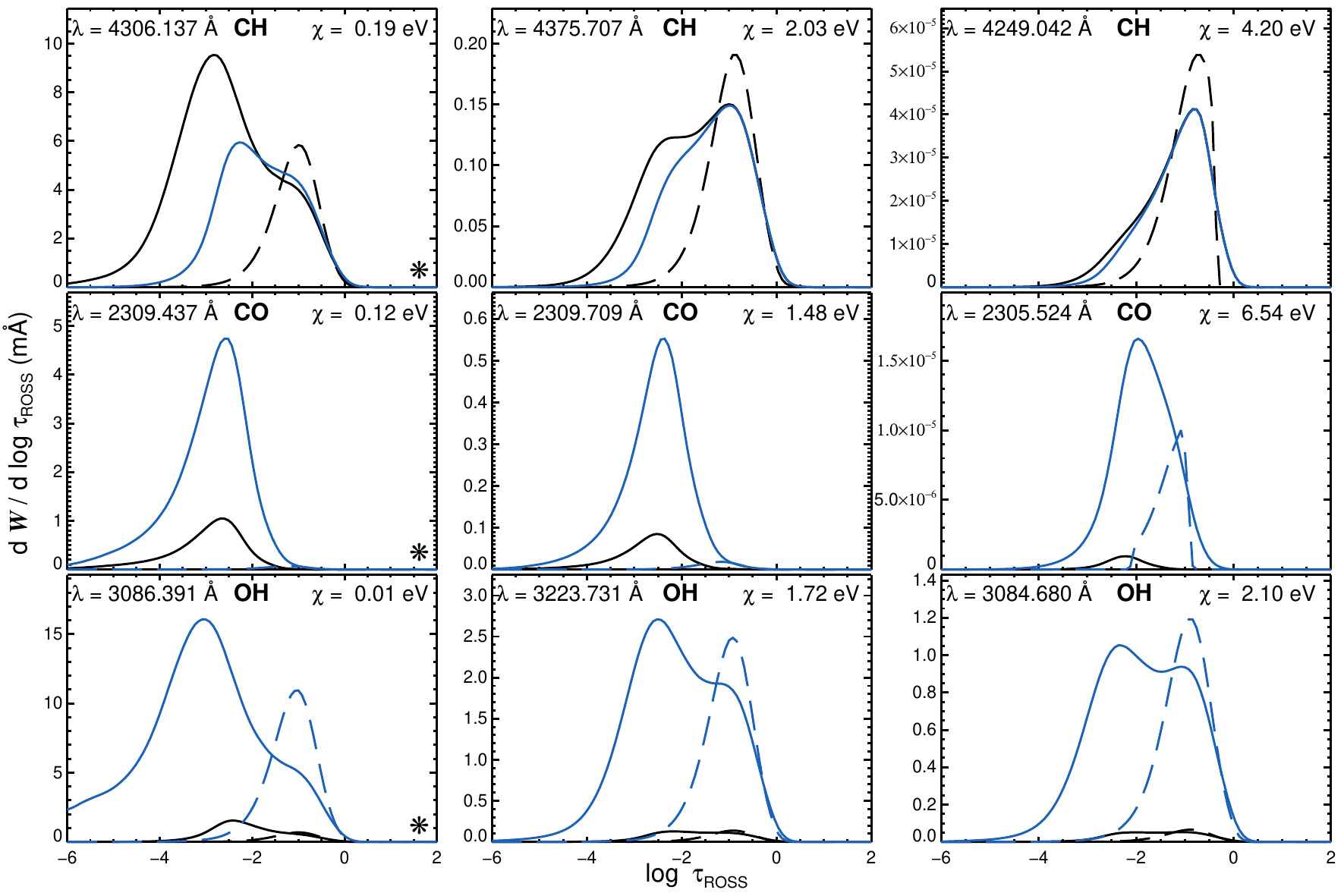}
	\caption{Corresponding $\feh=-2.0$ CH, CO and OH equivalent width contribution functions in 3D (solid lines) and and 1D (dashed lines) calculated for lines with different excitation potentials in the d3t63g40mm20n03 atmosphere. $\actd=\acod= 6.80\,{\rm dex}$ with $\aotd=\aood=6.20\,{\rm dex}$ (black) and $7.47\,{\rm dex}$ (blue). The oxygen abundance does not affect 1D CH contribution function.}
	\label{afig:contfunc2}
	\end{center}
\end{figure*}

\end{appendix}
\end{document}